%% file: fprv2011.tex
\documentclass[reprint,showpacs,superscriptaddress,preprintnumbers,amsmath,amssymb,aps,prd,nofootinbib]{revtex4-1}
\usepackage[utf8]{inputenc}
\usepackage{bbm}
\usepackage{hyperref}

\input{newcommands}

\begin{document}

\title{Nonperturbative semiclassical stability of de Sitter spacetime for small metric deviations}

\author{Markus B. Fröb}
\email{mfroeb@ffn.ub.edu}
\affiliation{Departament de Física Fonamental, Institut de Ciències del Cosmos (ICC), Universitat de Barcelona (UB), C/ Martí i Franquès 1, 08028 Barcelona, Spain}

\author{Demetrios B. Papadopoulos}
\email{papadop@astro.auth.gr}
\affiliation{Section of Astrophysics, Astronomy and Mechanics, Department of Physics, Aristotle University of Thessaloniki, 54124 Thessaloniki, Greece}

\author{Albert Roura}
\email{albert.roura@uni-ulm.de}
\affiliation{Max-Planck-Institut für Gravitationsphysik (Albert-Einstein-Institut), Am Mühlenberg 1, 14476 Golm, Germany}
\affiliation{Institut für Quantenphysik, Universität Ulm, Albert-Einstein-Allee 11, 89081 Ulm, Germany}

\author{Enric Verdaguer}
\email{enric.verdaguer@ub.edu}
\affiliation{Departament de Física Fonamental, Institut de Ciències del Cosmos (ICC), Universitat de Barcelona (UB), C/ Martí i Franquès 1, 08028 Barcelona, Spain}

\date{\today}
\begin{abstract}
We consider the linearized semiclassical Einstein equations for small deviations around de Sitter spacetime including the vacuum polarization effects of conformal fields. Employing the method of order reduction, we find the exact solutions for general metric perturbations (of scalar, vector and tensor type). Our exact (nonperturbative) solutions show clearly that in this case de Sitter is stable with respect to small metric deviations and a late-time attractor. Furthermore, they also reveal a breakdown of perturbative solutions for a sufficiently long evolution inside the horizon.
Our results are valid for any conformal theory, even self-interacting ones with arbitrarily strong coupling.
\end{abstract}


\maketitle

\section{Introduction}
\label{introduction}
\input{introduction}

\section{Semiclassical gravity and order reduction}
\label{semiclassical}

\subsection{Semiclassical Einstein equations}
\label{semiclassical_einstein}
\input{semiclassical_einstein}

\subsection{Order reduction}
\label{semiclassical_orderreduction}
\input{semiclassical_orderreduction}

\section{Conformal fields in a perturbed FLRW universe}
\label{model}

\subsection{The model}
\label{model_model}
\input{model_model}

\subsection{The FLRW background}
\label{model_flrw}
\input{model_flrw}

\section{Linear perturbations}
\label{linear}

\subsection{Gauge fixing}
\label{linear_gauge}
\input{linear_gauge}

\subsection{Semiclassical equations of motion}
\label{linear_eom}
\input{linear_eom}

\subsection{Nonlocal terms}
\label{linear_nonlocal}
\input{linear_nonlocal}

\section{Nonperturbative solutions}
\label{nonperturbative}

\subsection{Scalar and vector perturbations}
\label{nonperturbative_scalarvector}
\input{nonperturbative_scalarvector}

\subsection{Tensor perturbations}
\label{nonperturbative_tensor}
\input{nonperturbative_tensor}

\section{De Sitter stability and secular terms}
\label{stability_secular}

\subsection{Stability with respect to linear perturbations}
\label{stability_riemann}
\input{stability_riemann}

\subsection{Perturbative vs. nonperturbative solutions}
\label{stability_perturbative}
\input{stability_perturbative}

\section{Perturbed initial state}
\label{initial_states}
\input{initial_states}

\subsection{Tensor perturbations}
\label{initial_tensor}
\input{initial_tensor}

\subsection{Scalar and vector perturbations}
\label{initial_scalar_vector}
\input{initial_scalar_vector}

\section{General conformal field theories}
\label{general}
\input{general}

\section{Conclusions}
\label{conclusions}
\input{conclusions }

\begin{acknowledgments}
M.~F.\ acknowledges financial support through an APIF scholarship from the Universitat de Barcelona and a FPU scholarship no. AP2010-5453.
E.~V.\ and M.~F.\ also acknowledge partial financial support by the Research Projects MCI FPA2007-66665-C02-02, FPA2010-20807-C02-02, CPAN CSD2007-00042, with\-in the program Consolider-Ingenio 2010, and AGAUR 2009-SGR-00168. D. P. acknowledges financial support from the Aristotle University of Thessaloniki and the DAAD Foundation. He is grateful for the warm hospitality of the Department of Fundamental Physics of the UB and the Department of Theoretical Astrophysics of the University of Tübingen, where part of this work was carried out during his sabbatical leave.
During the final stages of this project, A.~R.\ was supported by the Deutsches Zentrum für Luft- und Raumfahrt (DLR) with funds provided by the Bundesministerium für Wirtschaft und Technologie under grant number DLR~50~WM~1136.
\end{acknowledgments}

\appendix

\section{Metric expansion}
\label{appendix_metric}
\input{appendix_metric}

\section{Conformal transformation}
\label{appendix_conformal}
\input{appendix_conformal}

\section{Special functions}
\label{appendix_special}
\input{appendix_special}

\section{Regular initial states}
\label{regular_states}
\input{regular_states}

\section{Comparison with Starobinsky's equation}
\label{appendix_starobinsky}
\input{appendix_starobinsky}

\bibliographystyle{apsrev4-1}
\bibliography{literature}

\end{document}

%% file: newcommands.tex
\newcommand{\mathe}{\mathrm{e}}
\newcommand{\mathi}{\mathrm{i}}
\newcommand{\mathpi}{\pi}
\let\oldre\Re
\let\oldim\Im
\renewcommand{\Re}{\oldre\mathfrak{e}\,}
\renewcommand{\Im}{\oldim\mathfrak{m}\,}

\newcommand{\dalembert}{\square}
\newcommand{\laplace}{\triangle}
\newcommand{\total}{\operatorname{d}\!}
\newcommand{\christoffel}[4][]{\,{}{#1}\Gamma^{#2}_{#3 #4}\,}
\newcommand{\varchristoffel}[4][]{\,{}{#1}\tilde\Gamma^{#2}_{#3 #4}\,}
\newcommand{\eqend}[1]{\,\text{#1}}
\newcommand{\bigo}[1]{\mathcal{O}\!\left({#1}\right)}

\newcommand{\Ein}{\operatorname{Ein}}
\DeclareMathOperator*{\distlim}{d-lim}
\newcommand{\sgn}{\operatorname{sgn}}
\newcommand{\abs}[1]{{\left\lvert{#1}\right\rvert}}
\newcommand{\bra}[1]{{\left\langle{#1}\right\rvert}}
\newcommand{\ket}[1]{{\left\lvert{#1}\right\rangle}}
\newcommand{\expect}[1]{{\left\langle{#1}\right\rangle}}

\newcommand{\op}[1]{\hat{#1}}

\renewcommand{\vec}[1]{\mathbf{#1}}
\newcommand{\const}{\mathrm{const.}}

%% file: introduction.tex
De Sitter space can be understood as an exponentially expanding cosmological spacetime entirely driven by a positive cosmological constant. It plays a central role in most models of cosmological inflation, where the potential-dominated energy density and pressure of the inflaton field act approximately as a cosmological constant and lead to a quasi-exponential accelerated expansion. The inflationary scenario provides a natural mechanism, through parametric amplification of quantum vacuum fluctuations, for generating a nearly scale-invariant spectrum of adiabatic primordial inhomogeneities, which can successfully explain the observed CMB anisotropies and the large scale structure of the universe \cite{mukhanov,lyth}.
Furthermore, the physics of de Sitter could also be important for elucidating the final fate of the universe if its current accelerated expansion is entirely due to a small cosmological constant, a possibility compatible with observations so far.

The exponential expansion quickly redshifts away any initial perturbations. This offers a simple means of establishing natural initial conditions for subsequent evolution once such an accelerated expansion has already started over a region with a size larger than the Hubble radius and lasts for a sufficiently large number of e-foldings. Under these conditions the initial classical perturbations are effectively erased and the quantum state for modes with wavelengths much smaller than the Hubble radius is very close to the Bunch-Davies or Euclidean vacuum, which locally is essentially equivalent to the Minkowski vacuum since at those length-scales the spacetime appears almost flat.
Such a scenario and the late-time attractor character of local de Sitter spacetime, often referred to as the ``no-hair'' property of de Sitter, is supported by a number of results and theorems in classical general relativity, both for linear perturbations \cite{barrow83,bruni01} as well as for the full nonlinear case \cite{wald83,starobinsky83,friedrich86,anderson05}.

It is, therefore, of great interest to establish whether those classical no-hair results can be extended to the quantum mechanical case. Solid conclusions have recently been obtained within the framework of quantum field theory in curved spacetime \cite{birrelldavies,waldqft}. Specifically, given a fairly general class of massive interacting theories with sufficiently weak coupling evolving on a fixed (nondynamical) de Sitter background, it has been shown to all orders in perturbation theory \cite{marolf10,hollands10} that quantum correlators within a spacetime region of bounded physical size become at sufficiently late times arbitrarily close to those of the Euclidean vacuum (the generalization of the de Sitter-invariant Bunch-Davies vacuum to interacting theories).

Considering test fields evolving on a fixed background, however, offers an incomplete answer: addressing the full dynamical problem requires taking into account the back-reaction of the quantum fields on the dynamics of the spacetime geometry. A number of studies have explored this question in the context of semiclassical gravity, where the metric is still treated classically, but its dynamics is governed by a generalization of the Einstein equation which includes the expectation value of the stress tensor operator of the quantum matter fields as a source \cite{waldqft,flanaganwald}.
Focusing on the backreaction of conformal matter fields, the dynamics of scalar-type metric perturbations around de Sitter has been analyzed in ref.~\cite{anderson09}, where the importance of considering also perturbations of the initial state of the matter fields has been emphasized. (The linear stability of de Sitter including initial classical stress tensor sources had earlier been studied in refs.~\cite{ginsparg83,frieman82}.)
In addition, the evolution of tensor metric perturbations for the same situation has been studied in ref.~\cite{fordetal}, where the semiclassical Einstein equation was solved perturbatively. (As we will show below, however, these perturbative solutions cease to be valid for a sufficiently long evolution inside the horizon.)
The stability of tensor perturbations including the backreaction from conformal fields has also been considered \cite{pelinson12} in investigations on the existence of unstable runaway solutions of the corresponding higher-order equations in the context of inflationary models driven by the trace anomaly \cite{starobinskyinflation}. Nevertheless, those analysis neglected the contribution of nonlocal terms which play a key role in the existence of runaway solution for perturbations around flat space \cite{horowitz80,flanaganwald}.

Related studies have also been carried out for nonconformal fields in de Sitter spacetime. The case of massless minimally coupled free scalar fields has been considered in ref.~\cite{park11}
and a vanishing correction to the classical modes was found when solving perturbatively the linearized semiclassical equation for tensor perturbations. Both massless and massive nonconformal scalar fields were studied in refs.~\cite{pereznadal08a,pereznadal08b} and the stability of de Sitter spacetime with respect to spatially isotropic perturbations was established. A fairly general class of Gaussian initial states was considered and de Sitter was found to be a late-time attractor in all cases. Moreover, the importance of taking into account the contribution of nonlocal terms for light massive fields when analyzing the stability in the infrared regime was elucidated~\cite{pereznadal08b}.

Here we consider the linearized semiclassical Einstein equation around a de Sitter background including the vacuum polarization effects of conformal matter fields, and solve it exactly for general metric perturbations (of scalar, vector and tensor type).
In doing so, we make use of the method of order reduction \cite{simonparker93,flanaganwald}, which eliminates the spurious solutions associated with higher-order derivatives while capturing the right dynamics in the infrared regime. Moreover, the method generates a backreaction equation which is equivalent to the original semiclassical Einstein equation up to the same order in inverse powers of the Planck mass at which the latter is valid within an effective field theory (EFT) approach to perturbative quantum gravity \cite{donoghue94,burgess03}, but which can be significantly simpler to solve, a fact that we exploit in our calculation.
Our exact solutions clearly show that de Sitter spacetime is also stable in this case and a late-time attractor as far as local geometrical properties are concerned. Furthermore, it reveals a breakdown of perturbation theory when solving the semiclassical equation for a long time evolution inside the horizon.

It should be stressed that our results are valid for any conformal field theory (CFT), even self-interacting ones with arbitrary strong coupling, as explained in sec.~\ref{conclusions}. In addition to metric perturbations, perturbed initial states of the matter fields have also been considered.

Although semiclassical gravity does take into account the backreaction of the quantum matter fields on the dynamics of the mean spacetime geometry, it does not provide a complete analysis because it does not include the quantum mechanical effects of the metric itself. Indeed, one needs to quantize the metric perturbations in order to account for certain relevant phenomena: doing so is necessary, for instance, for a proper description of the generation of primordial cosmological perturbations, and it has even been suggested that radiative corrections involving higher-order graviton loops could lead to a secular screening of the cosmological constant \cite{tsamis96,tsamis97}. However, detailed calculations including graviton loops are technically complex and one is, in addition, confronted with the need to consider appropriate observables which are not only gauge-invariant 
beyond linear order but also infrared safe \cite{gerstenlauer11,giddings11,urakawa10}. Because of such difficulties only partial progress has been made in this direction. It is, therefore, important to consider also somewhat less ambitious problems, but obtain solid results (and, if possible, exact) which can provide a robust foundation for further developments.
One such example is the exact calculation of one-loop corrections from matter fields to the correlator of the Riemann tensor \cite{pereznadal10,frv2011a,frv2012} for quantized metric perturbations around de Sitter. The results support the existence of quantum states for metric perturbations interacting with matter fields which exhibit (appropriately defined) de Sitter invariance, at least when graviton loops are neglected.
In contrast, the results presented here only apply to the mean field geometry, but explore the effect of different (non-de Sitter-invariant) initial states on the dynamics and show not only that a self-consistent de Sitter-invariant solution exists, but also that it is a late-time attractor.
Furthermore, the methods described below for obtaining nonperturbative solutions valid for long evolution times could prove helpful in order to extend the calculation of the Riemann correlator so that it correctly captures the details of its behavior for large separations (both spatial and temporal), which seems to require a nonperturbative treatment.

The rest of the paper is organized as follows.
Semiclassical gravity and the method of order reduction are briefly reviewed in sec.~\ref{semiclassical}. The semiclassical Einstein equation for metric perturbations around a spatially flat FLRW spacetime including the quantum back-reaction of conformal fields is presented in sec.~\ref{model}. Given a cosmological constant and fields in the Bunch-Davies vacuum, there is a self-consistent semiclassical de Sitter background.
In sec.~\ref{linear} the metric perturbations are decomposed into scalar, vector and tensor contributions. Next, we fix the gauge and write the (decoupled) semiclassical equations for the three types of perturbations, before and after employing the order reduction method. The exact (nonperturbative) solutions are obtained in sec.~\ref{nonperturbative} and their implications for the stability of de Sitter spacetime as well as the breakdown of the perturbative solutions are analyzed in sec.~\ref{stability_secular}.
The effects of perturbing also the initial state of the matter fields are studied in sec.~\ref{initial_states} and we show that all our main conclusions remain unchanged.
Finally, in sec.~\ref{conclusions} we summarize and discuss our results as well as explaining their applicability to any CFT for the matter fields.
Several useful formulae concerning the perturbative expansion of curvature tensors, their conformal transformations and some special functions are provided in the first three appendices.
In addition, a method for generating a family of regular Gaussian initial states is described in appendix~\ref{regular_states}, and in appendix~\ref{appendix_starobinsky} we compare our linearized semiclassical Einstein equation for tensor perturbations (before order reduction) with the one previously obtained by Starobinsky \cite{starobinsky81}.

Throughout the paper we use natural units with $c = \hbar = 1$ and take $\kappa^2 = 16 \mathpi G_\text{N}$. We employ the ``+++'' sign convention of ref.~\cite{mtw} and Greek indices range over space and time, while Latin indices denote spatial components only.

%% file: semiclassical_einstein.tex
Semiclassical gravity can be regarded as a mean field approximation to a quantum theory of gravity where the mean gravitational field is treated classically and only matter is quantized. In contrast to quantum field theory in curved spacetime, it also includes the back-reaction of the matter fields on the mean geometry, given by the background metric $g_{\mu\nu}$. To achieve this, the stress tensor on the right-hand side of the Einstein equation is replaced by the expectation value of an appropriate quantum stress tensor operator. This expectation value needs to be renormalized, and counterterms local in the gravitational field have to be included in the bare gravitational action. At all loop order for the matter fields (but no graviton loops) those counterterms are quadratic in the curvature and the renormalized semiclassical Einstein equation reads
\begin{equation}
\label{einstein_semiclassical}
\begin{split}
G_{\mu\nu} + \Lambda(\mu) g_{\mu\nu} =& \ a_1(\mu) A_{\mu\nu} + a_2(\mu) B_{\mu\nu} \\
&+ \frac{1}{2} \kappa^2(\mu) \expect{\op{T}_{\mu\nu}(\mu)}_\text{ren} \eqend{,}
\end{split}
\end{equation}
where $A_{\mu\nu} = (-g)^{-1/2} \, (\delta/\delta g^{\mu\nu}) \int C^{\alpha\beta\gamma\delta} C_{\alpha\beta\gamma\delta} \sqrt{-g} \total^4 x$ and $B_{\mu\nu} = (-g)^{-1/2} \, (\delta/\delta g^{\mu\nu}) \int R^2 \sqrt{-g} \total^4 x$ are obtained by functionally differentiating the finite parts of the gravitational counterterms. The parameters $a_1(\mu)$ and $a_2(\mu)$ together with $\Lambda(\mu)$ and $\kappa^2(\mu)$ are in general renormalized parameters which have to be determined by experiment, and $\mu$ is the renormalization scale.
Note, nevertheless, that the backreaction equation~\eqref{einstein_semiclassical} is renormalization group invariant and the dependence on $\mu$ of the different parameters appearing in the equation and the renormalized expectation value $\langle \op{T}_{\mu\nu}(\mu) \rangle_\text{ren}$ cancel out.

The tensors $A_{\mu\nu}$ and $B_{\mu\nu}$ are explicitly given by
\begin{equation}
\label{tensors_ab_definition}
\begin{split}
A_{\mu\nu} &= - 4 \nabla^{(\alpha} \nabla^{\beta)} C_{\alpha\nu\beta\mu} - 2 R^{\alpha\beta} C_{\alpha\mu\beta\nu} \\
&= - 4 R_{\alpha\mu} R^\alpha_\nu + \frac{4}{3} R R_{\mu\nu} - \frac{1}{3} g_{\mu\nu} R^2 + g_{\mu\nu} R^{\alpha\beta} R_{\alpha\beta} \\
&\quad+ 8 \nabla_{[\alpha} \nabla_{\mu]} R^\alpha_\nu - 2 \dalembert_g R_{\mu\nu} + \frac{2}{3} \nabla_\mu \nabla_\nu R + \frac{1}{3} g_{\mu\nu} \dalembert_g R \\
B_{\mu\nu} &= \frac{1}{2} g_{\mu\nu} R^2 - 2 R R_{\mu\nu} + 2 \nabla_\mu \nabla_\nu R - 2 g_{\mu\nu} \dalembert_g R \eqend{,}
\end{split}
\end{equation}
where the two forms of $A_{\mu\nu}$ given above were obtained by using the identity \eqref{fourd_weyl_identity} and the definition of the Weyl tensor \eqref{weyl_tensor_definition} as well as the second Bianchi identity.

%% file: semiclassical_orderreduction.tex
The semiclassical Einstein equation~\eqref{einstein_semiclassical} contains terms with up to fourth-order derivatives of the metric, as seen from eq.~\eqref{tensors_ab_definition}. (The expectation value $\langle \op{T}_{\mu\nu}(\mu) \rangle_\text{ren}$ also involves similar terms, as shown below.)
Such kind of higher-order time derivatives are common in backreaction problems. A well known example is the Abraham-Lorentz-Dirac equation, which describes the effect of radiation reaction on the motion of a point-like charge in classical electrodynamics \cite{jackson,lifshitzbook} (i.e.\ without considering the internal structure of the particle nor a finite size for the charge density distribution). In fact, they are a generic feature of effective field theories (EFTs), where the effects of the UV sector on the dynamics of the low-energy degrees of freedom are encoded at the level of the action through an expansion of local terms with an increasing number of derivatives.
The validity of the EFT expansion relies on the fact that for length-scales much larger than the inverse cut-off scale of the UV sector the higher-order terms in the expansion become increasingly smaller. In this regime their contribution amounts to a small correction to the equation of motion which results, when treated perturbatively, into locally small perturbations of the classical solutions.
In contrast, solving the corresponding higher-order equations exactly gives rise to additional solutions exhibiting exponential instabilities with characteristic time-scales comparable to the inverse cutoff scale of the EFT (or sometimes fast oscillations with the same kind of characteristic timescale), often referred to as ``runaway'' solutions. These are spurious solutions which should not be taken seriously since they involve characteristic scales for which the EFT expansion breaks down and the contributions from the higher-order terms to the equation of motion no longer correspond to small corrections but to dominant terms.

The simplest way of avoiding such spurious solutions is by solving the corrected equations of motion perturbatively. However, perturbative solutions may not be valid for long times. This happens when quantities like the total time appear multiplying the perturbative parameter so that the expansion contains so-called secular terms which grow with time and lead to a breakdown for sufficiently long times of the truncated perturbative expansion.
Those limitations can be overcome with the \emph{order reduction} method, which consists in taking the equation of motion with corrections up to a finite order and writing an alternative equation which is equivalent up to that order but contains no higher derivative terms
(this is achieved by taking successive derivatives of the original equation and substituting the higher-order derivatives in the correction terms to the appropriate order).
The exact solutions of the equation obtained with this method agree locally with the perturbative solutions constructed around different times (each one with a finite domain of validity) and provides an interpolation between all of them valid for long times.
This is particularly important when considering situations where the effects of the corrections are locally small, but can build up over long times and give rise to substantial accumulated effects. Two examples of such situations are an electric charge following a quasi-circular trajectory in a uniform magnetic field
and emitting electromagnetic radiation for a sufficiently long time so that the radius of its orbit decreases, say, to half of its initial value due to radiation reaction, or an evaporating black hole emitting Hawking radiation for such a long time that its mass (or horizon size) decreases to a small fraction of its initial value.

Related alternative methods which have been employed in the literature for discarding the spurious solutions mentioned above involve finding the exact solutions of the original backreaction equation and then selecting the appropriate subset either by demanding analyticity of the solutions with respect to the perturbative parameter or checking explicitly which solutions exhibit unphysical characteristic scales and disregarding them. However, the latter method is less systematic and requires a case-by-case analysis, whereas the analyticity requirement may be too restrictive in some cases \cite{flanaganwald}.
Furthermore, the order reduction method leads to equations of motion which are equivalent up to the order under consideration, but are often easier to solve, as will be the case for the problem analyzed in the remaining sections.

The order reduction method has been applied to electromagnetic \cite{lifshitzbook} and gravitational \cite{beletal81} radiation reaction problems as well as higher derivative gravity \cite{belzia85}. It has also been employed in semiclassical gravity \cite{simonparker93,flanaganwald} and in this context it has been argued \cite{simon92} that trace-anomaly-driven inflationary models (with no cosmological constant and driven entirely by the vacuum polarization of large number of matter fields \cite{starobinskyinflation}) correspond to spurious solutions which lie beyond the EFT's domain of applicability and are automatically discarded when using order reduction.

It should be noted that order reduction cannot be always applied in a straightforward way. It may be ambiguous in integro-differential equations, or may lead to covariance breaking if the time derivatives and spatial derivatives are not simultaneously reduced; see \cite{flanaganwald} for a detailed discussion of these issues.

The order reduction method can be illustrated in a nutshell with the following simple example of a first order differential equation in time for a function $f(\eta)$ with a perturbative correction of order $\kappa^2$. Given
\begin{equation}
\label{order_reduction_eq}
f' + b f = \kappa^2 P(f, f', f'', ...) \eqend{,}
\end{equation}
where $b$ is a constant and $P$ is an arbitrary function, order reduction uses that $f' = - b f + \bigo{\kappa^2}$, and by deriving one more time $f'' = - b f' + \bigo{\kappa^2} = b^2 f + \bigo{\kappa^2}$. Substituting those two equations into the right hand side, we get
\begin{equation}
\label{order_reduction_eq_reduced}
f' + b f = \kappa^2 P(f, - b f, b^2 f, ...) + \bigo{\kappa^4} \eqend{,}
\end{equation}
which is an equation of first order which is valid to the same order in $\kappa^2$ as the original equation \eqref{order_reduction_eq}, but does not have unphysical solutions.
Rather than considering a truncated perturbative expansion, this equation can now be solved exactly.
It is clear how the method works for equations of more derivatives or partial differential equations: one takes the lowest order equation and substitutes it in the higher order terms (in $\kappa^2$), taking additional derivatives if necessary.

%% file: model_model.tex
Our model consists of $N$ massless free scalar fields conformally coupled to the spacetime curvature:
\begin{equation}
\label{field_action}
S[\varphi,\tilde{g}] = -\frac{1}{2} \sum_{j=1}^N \int \left[ \tilde{g}^{\mu\nu}
\partial_\mu \varphi_j \, \partial_\nu \varphi_j  + \xi_\text{cc} \tilde{R}\, \varphi_j^2 \right]
\sqrt{-\tilde{g}} \total^d x
\eqend{,}
\end{equation}
where $\tilde{R}$ is the Ricci scalar associated with the metric $\tilde{g}_{\mu\nu}$ and $\xi_\text{cc} = (d-2)/4(d-1)$, which reduces to $\xi_\text{cc} = 1/6$ in four dimensions. We will use dimensional regularization, but after the renormalization procedure has been carried out we will take $d=4$.
Furthermore, we will specialize the physical metric $\tilde{g}_{\mu\nu}$ to a slightly perturbed spatially flat FLRW spacetime, which is conformal to an almost flat metric $g_{\mu\nu}$:
\begin{equation}
\label{metric_perturbation}
\tilde{g}_{\mu\nu} = a^2(\eta) g_{\mu\nu} = a^2(\eta) ( \eta_{\mu\nu} + h_{\mu\nu} ) \eqend{,}
\end{equation}
where $\eta$ denotes the conformal time.

The renormalized semiclassical Einstein equation \eqref{einstein_semiclassical} for this model with the fields in the conformal vacuum state was derived by Campos and Verdaguer in refs.~\cite{camposverdaguer94,camposverdaguer96} using the closed-time-path (CTP) effective action, and is given to linear order in the perturbation $h_{\mu\nu}$ by
\begin{widetext}
\begin{equation}
\label{einstein_semiclassical_conformal}
\begin{split}
\tilde{G}_{\mu\nu} + \Lambda \tilde{g}_{\mu\nu} &= - \frac{1}{12} \beta \kappa^2 \tilde{B}_{\mu\nu} + \frac{1}{2} \alpha \kappa^2 \left[ \tilde{H}_{\mu\nu} - 2 \tilde{R}^{\alpha\beta} \tilde{C}_{\mu\alpha\nu\beta} \right] \\
&\quad\qquad+ \frac{3}{2} \alpha \kappa^2 a^{-2} \left[ - 4 \nabla^\alpha \nabla^\beta \left( C_{\mu\alpha\nu\beta} \ln a \right) + \int H(x-y; \bar{\mu})\, A_{\mu\nu}(y) \total^4 y \right] \eqend{.}
\end{split}
\end{equation}
\end{widetext}
where $\alpha = N/(2880 \mathpi^2)$.
Here and throughout the rest of the paper $\nabla_\mu$ denotes the covariant derivative with respect to the metric $g_{\mu\nu}$ and $\dalembert_g = \nabla^\mu \nabla_\mu$, whereas objects with a tilde are evaluated with the conformally related metric $\tilde{g}_{\mu\nu}$. Furthermore, background quantities will be denoted by a superscript ${}^{(0)}$ such as $g^{(0)}_{\mu\nu} = a^2(\eta) \eta_{\mu\nu}$, and quantities linearized in the perturbation by a superscript ${}^{(1)}$ as in $\tilde{g}^{(1)}_{\mu\nu} = a^2(\eta) h_{\mu\nu}$.

For conformal fields the renormalized parameter $a_2(\mu)$ does not depend on $\mu$ and we denote it by $\beta$. Moreover we have chosen a renormalization scale $\bar{\mu}$ such that $a_1(\bar{\mu}) = 0$.
Similarly, both $\Lambda$ and $\kappa^2$ are also independent of $\mu$ in this case. The tensors $A_{\mu\nu}$ and $B_{\mu\nu}$ are defined in equation~\eqref{tensors_ab_definition} and
\begin{equation}
\label{tensors_definition}
\begin{split}
H_{\mu\nu} &= - R_{\mu\sigma} {R^\sigma}_\nu + \frac{2}{3} R R_{\mu\nu} + \frac{1}{2} g_{\mu\nu} R^{\alpha\beta} R_{\alpha\beta} - \frac{1}{4} g_{\mu\nu} R^2 \eqend{.}
\end{split}
\end{equation}
Finally, the kernel $H(x-y; \mu)$ depending on the renormalization scale $\mu$ is given in appendix \ref{appendix_starobinsky}, but its exact form will not be needed in the bulk of the paper. Eq.~\eqref{einstein_semiclassical_conformal} coincides with those derived by alternative methods \cite{horowitzwald1,horowitzwald2,starobinsky81}.

%% file: model_flrw.tex
The semiclassical generalization of the Friedmann equation can be obtained by setting the perturbation $h_{\mu\nu}$ to zero in the $00$ component of eq.~\eqref{einstein_semiclassical_conformal}, which gives
\begin{equation}
\label{friedmann_equation}
\begin{split}
6 (a')^2 - 2 \Lambda a^4 &= 3 \alpha \kappa^2 a^{-4} (a')^4 \\
&+ 3 \beta \kappa^2 a^{-3} \left[ 2 a a' a''' - a (a'')^2 - 4 (a')^2 a'' \right] \eqend{,}
\end{split}
\end{equation}
and using the order reduction method it becomes
\begin{equation}
\label{friedmann_equation_orderreduction}
\left( a' \right)^2 - \frac{\Lambda}{3} \left[ 1 + \frac{1}{6} \alpha \kappa^2 \Lambda \right] a^4 = \bigo{\kappa^4} \eqend{.}
\end{equation}
Defining an effective cosmological constant $\Lambda_\text{eff}$ as
\begin{equation}
\label{def_lambda_eff}
\Lambda_\text{eff} = \Lambda \left[ 1 + \frac{1}{6} \alpha \kappa^2 \Lambda \right] \eqend{,}
\end{equation}
Eq.~\eqref{friedmann_equation_orderreduction} has the solution
\begin{equation}
\label{scale_factor}
a(\eta) = - \frac{1}{H \eta} \eqend{,}
\end{equation}
where the Hubble parameter $H$ is given by $3 H^2 = \Lambda_\text{eff}$, and $-\infty < \eta \leq 0$. This solution is unique up to a shift of the origin of conformal time, $\eta \to \eta - \eta_0$, and its sign.

Hence, de Sitter spacetime, given here in spatially flat coordinates (the Poincaré patch), is a self-consistent solution of the semiclassical Friedmann equation~\eqref{friedmann_equation}, with the effective cosmological constant~\eqref{def_lambda_eff} having a small positive shift of quantum origin.
The existence of such self-consistent solutions follows straightforwardly from the fact that the renormalized expectation value of the stress tensor for the Bunch-Davies vacuum must be proportional to the metric, as implied by de Sitter invariance, and has been know for a long time \cite{dowker76,wada83}.
When $\Lambda = 0$, eq.~\eqref{friedmann_equation} still admits a de Sitter solution with $H^2 = 2/(\alpha \kappa^2)$ (closely connected to Starobinsky's original model of inflation \cite{starobinskyinflation,vilenkin85}), but its characteristic scale lies beyond the domain of validity of semiclassical gravity when regarded as part of an EFT approach to quantum gravity, as briefly discussed in sec.~\ref{semiclassical_orderreduction}. Such solutions are automatically discarded by the method of order reduction \cite{simonparker93}.

%% file: linear_gauge.tex
When considering metric perturbations around a given background, there is a gauge freedom (corresponding to local diffeomorphisms) associated with the mapping between the background and the perturbed geometry.
Infinitessimal diffeomorphisms generated by an arbitrary vector field $\tilde{\xi}^\nu$ induce the following gauge transformation of the physical perturbation $a^2 h_{\mu\nu}$ introduced in Eq.~\eqref{metric_perturbation}: 
\begin{equation}
\label{h_decomp}
a^2 h_{\mu\nu} \to a^2 h_{\mu\nu} + \tilde{\nabla}_\mu \tilde{\xi}_\nu + \tilde{\nabla}_\nu \tilde{\xi}_\mu \eqend{,}
\end{equation}
Rescaling the arbitrary vector field $\tilde{\xi}_\mu = a^2 \xi_\mu$ and using Eq.~\eqref{scale_factor} for the scale factor $a(\eta)$, Eq.~\eqref{h_decomp} becomes
\begin{equation}
\label{gauge_trafo}
h_{\mu\nu} \to h_{\mu\nu} + \partial_\mu \xi_\nu + \partial_\nu \xi_\mu + \frac{2}{\eta} \eta_{\mu\nu} \xi_0 \eqend{.}
\end{equation}
Before proceeding any further, it is convenient to decompose the perturbation $h_{\mu\nu}$
exploiting the fact that the spatial sections of the background metric are maximally symmetric spaces \cite{lifshitz,bardeen,stewart90}.
The spatial part $h_{ij}$ decomposes as
\begin{equation}
\label{perturbation_spatial_decompose}
h_{ij} = h^\text{TT}_{ij} + 2 \partial_{(i} w^\text{T}_{j)} + \partial_i \partial_j \sigma + \tau \delta_{ij} \eqend{,}
\end{equation}
where $\delta^{ij} \partial_i h^\text{TT}_{jk} = 0 = \delta^{ij} h^\text{TT}_{ij}$ and $\delta^{ij} \partial_i w^\text{T}_j = 0$. The temporal components can be similarly decomposed as
\begin{equation}
\label{perturbation_temporal_decompose}
h_{0i} = v^\text{T}_i + \partial_i \psi \eqend{,} \qquad h_{00} = \phi \eqend{,}
\end{equation}
where $\delta^{ij} \partial_i v^\text{T}_j = 0$. In total, we have four scalars $\phi$, $\psi$, $\sigma$, $\tau$, two transverse vectors $w^\text{T}_i$ and $v^\text{T}_i$ (with two independent components each) and a transverse traceless tensor $h^\text{TT}_{ij}$ (with two independent components as well). 
Decomposing also the spatial part of the vector field $\xi_\mu$ as
\begin{equation}
\xi_i = \xi^\text{T}_i + \partial_i \xi \eqend{,}
\end{equation}
where $\delta^{ij} \partial_i \xi^\text{T}_j = 0$, we can see the behavior of the various components under a gauge transformation:
\begin{equation}
\begin{aligned}
h^\text{TT}_{ij} &\to h^\text{TT}_{ij} & w^\text{T}_i &\to w^\text{T}_i + \xi^\text{T}_i \\
\sigma &\to \sigma + 2 \xi & \tau &\to \tau + \frac{2}{\eta} \xi_0 \\
v^\text{T}_i &\to v^\text{T}_i + \xi^{\prime\text{T}}_i & \psi &\to \psi + \xi' + \xi_0 \\
\phi &\to \phi + 2 \xi'_0 - \frac{2}{\eta} \xi_0 \eqend{,} & &
\end{aligned}
\end{equation}
where primes denotes derivatives with respect to the conformal time $\eta$.
Choosing $\xi^\text{T}_i$, $\xi$ and $\xi_0$ appropriately, we can set $w^\text{T}_i$, $\sigma$ and $\tau$ to zero. We will work in this gauge, where the perturbation of the spatial metric is entirely given by the tensorial component:
\begin{equation}
\label{TTgauge}
h_{ij} = h^\text{TT}_{ij}.
\end{equation}
This is the \emph{transverse traceless} gauge, also known as \emph{spatially flat} gauge when focusing on the scalar perturbations.
This fixes completely the gauge if we restrict ourselves to metric perturbations that fall off at spatial infinity.

If we consider perturbations which do not necessarily fall off, there is still some residual gauge freedom which is not fixed by condition \eqref{TTgauge}.
On one hand, there are transformations which leave $h_{ij}$ invariant. One possibility are those generated by $\xi^\text{T}_i$ which are functions only of the conformal time; this corresponds to translations on the spatial sections which can change for each surface of the foliation and lead to changes of $v^\text{T}_i$.
A second possibility involves transformations generated by $\xi_i = b(\eta) \delta_{ij} x^j$, $\xi_0 = - b(\eta) \eta$; this corresponds to dilations on the spatial sections while changing at the same time the surface of the foliation so that the expansion of the FLRW background compensates for that, and leads to changes of $\phi$ and $\psi$ while leaving $h_{ij}$ invariant.
On the other hand, the transformations generated by $\xi^\text{T}_i = E_{ij} x^j$ with $E_{ij}$ constant and traceless induce changes of $h_{ij}$ but leave it transverse and traceless. 
These residual gauge transformations will play a role in sec.~\ref{nonperturbative} to show that certain solutions are pure gauge.

%% file: linear_eom.tex
Using the metric decomposition and the gauge fixing introduced in the previous subsection, one can obtain from the semiclassical equation~\eqref{einstein_semiclassical_conformal} the dynamical equations for the scalar, vector and tensor perturbations, which decouple form each other.
From the $00$ component of eq.~\eqref{einstein_semiclassical_conformal} and the scalar part of its $0i$ component, one gets the following two equations for the scalar perturbations $\psi$ and $\phi$:
\begin{widetext}
\begin{equation}
\label{semiclassical_scalar}
\begin{split}
0 &= - \left( 3 - 5 \alpha \kappa^2 H^2 \right) H^2 a^4 \phi + 2 \left[ 1 - \left( 3 \alpha - \beta \right) \kappa^2 H^2 \right] H a^3 \laplace \psi + \frac{1}{2} \beta \kappa^2 \bigg[ - 7 H^2 a^2 \laplace \phi + 3 H^2 a^2 \phi'' + 6 H^3 a^3 \phi' - \frac{1}{3} \laplace^2 \phi \bigg] \\
&\quad+ \beta \kappa^2 \bigg[ - H a \laplace \psi'' + H a \laplace^2 \psi + \frac{1}{3} \laplace^2 \psi' \bigg] - \alpha \kappa^2 \int \bigg[ 2 \laplace^2 \psi'(x') - \laplace^2 \phi(x') \bigg] \left( H(x-x'; \bar{\mu}) + \delta^4(x-x') \ln a \right) \total^4 x' ,\\
0 &= 2 H a^3 \laplace \phi + 2 \alpha \kappa^2 \left( H a \laplace^2 \phi - 3 H^3 a^3 \laplace \phi - 2 H a \laplace^2 \psi' \right) + \beta \kappa^2 \left( H a \dalembert \laplace \phi - 2 H^2 a^2 \laplace \phi' - \frac{1}{3} \laplace^2 \phi' \right) \\
&\quad+ \beta \kappa^2 \left( - 4 H^2 a^2 \laplace^2 \psi + \frac{2}{3} \laplace^2 \psi'' \right) - 2 \alpha \kappa^2 \int \bigg[ 2 \laplace^2 \psi''(x') - \laplace^2 \phi'(x') \bigg] \left( H(x-x'; \bar{\mu}) + \delta^4(x-x') \ln a \right) \total^4 x' ,\\
\end{split}
\end{equation}
where $\dalembert = \eta^{\mu\nu} \partial_\mu \partial_\nu$ and $\laplace = \delta^{ij} \partial_i \partial_j$.
Similarly, from the transverse part of the $0i$ component one gets the equation for the vector perturbation $v^\text{T}_i$:
\begin{equation}
\label{semiclassical_vector}
0 = \left[ 1 - \left( \alpha + 2 \beta \right) \kappa^2 H^2 \right] a^2 \laplace v^\text{T}_i - 3 \alpha \kappa^2 H a \laplace v^{\prime\text{T}}_i + 3 \alpha \kappa^2 \int \left( \dalembert \laplace v^\text{T}_i(x') \right) \left( H(x-x'; \bar{\mu}) + \delta^4(x-x') \ln a \right) \total^4 x' .
\end{equation}
Finally, the equation for the tensor perturbations is obtained from the transverse and traceless part of the $ij$ components:
\begin{equation}
\label{semiclassical_tensor}
\begin{split}
0 &= - 2 \left[ 1 - \left( \alpha + 2 \beta \right) \kappa^2 H^2 \right] H a^3 h^{\prime\text{TT}}_{ij} + \left[ 1 - \left( \alpha + 2 \beta \right) \kappa^2 H^2 \right] a^2 \dalembert h^\text{TT}_{ij} + 3 \alpha \kappa^2 H^2 a^2 \left[ 2 h^{\prime\prime\text{TT}}_{ij} + \dalembert h^\text{TT}_{ij} \right] \\
&\quad- 6 \alpha \kappa^2 H a \dalembert h^{\prime\text{TT}}_{ij} + 3 \alpha \kappa^2 \int \left( \dalembert \dalembert h^\text{TT}_{ij}(x') \right) \left( H(x-x'; \bar{\mu}) + \delta^4(x-x') \ln a \right) \total^4 x' .
\end{split}
\end{equation}
\end{widetext}

Employing order reduction as explained in sec.~\ref{semiclassical_orderreduction}, the equations can be rewritten in the much simpler form
\clearpage
\begin{subequations}
\label{semiclassical_perturbation_or}
\begin{align}
h^{\prime\prime \text{TT}}_{ij} - \frac{2}{\eta} \left( 1 - \nu \right) h^{\prime \text{TT}}_{ij} - \left( 1 - 2 \nu \right) \laplace h^\text{TT}_{ij} &= \bigo{\kappa^4} \label{semiclassical_perturbation_or_tensor} \eqend{,} \\
\laplace v^\text{T}_i &= \bigo{\kappa^4} \label{semiclassical_perturbation_or_vector} \eqend{,} \\
\laplace \phi &= \bigo{\kappa^4} \label{semiclassical_perturbation_or_phi} \eqend{,} \\
\laplace \psi + \frac{3}{2 \eta} \left( 1 + \frac{4}{9} \nu \right) \phi &= \bigo{\kappa^4} \label{semiclassical_perturbation_or_psi} \eqend{,}
\end{align}
\end{subequations}
where we have introduced the following parameter, which controls the expansion in powers of $\kappa^2$:
\begin{equation}
\label{def_nu}
\nu = 3 \alpha \kappa^2 H^2 \ll 1 \eqend{.}
\end{equation}
It is worth emphasizing that those equations are independent of the arbitrary parameter $\beta$ and the renormalization scale $\bar{\mu}$ of the semiclassical theory. They involve only the semiclassical parameter $\alpha$, which depends on the matter field content.

%% file: linear_nonlocal.tex
As we have calculated explicitly, when we use order redution the nonlocal terms do not contribute to the semiclassical equations of motion~\eqref{semiclassical_perturbation_or} for the perturbation $h_{\mu\nu}$. We now want to show that one can see this in general, without choosing a gauge or expanding the semiclassical equations explicitly in terms of the perturbation $h_{\mu\nu}$. Basically this amounts to showing that $A_{\mu\nu}$, given by eq.~\eqref{tensors_ab_definition}, is of order $\kappa^2$ when using order reduction. In order to do so, it is convenient to consider its definition as a functional derivative of the integral of the square of the Weyl tensor
\begin{equation}
A_{\mu\nu} = \frac{1}{\sqrt{-g}} \frac{\delta F}{\delta g^{\mu\nu}}
\end{equation}
with
\begin{equation}
F = \int C^{\alpha\beta\gamma\delta} C_{\alpha\beta\gamma\delta} \sqrt{-g} \total^4 x = \int \tilde{C}^{\alpha\beta\gamma\delta} \tilde{C}_{\alpha\beta\gamma\delta} \sqrt{-\tilde{g}} \total^4 x \eqend{,}
\end{equation}
where the last equality follows from the conformal invariance of the Weyl tensor with one raised index. This also implies that
\begin{equation}
\label{amunu_atildemunu}
A_{\mu\nu} = \frac{1}{\sqrt{-g}} \frac{\delta F}{\, \delta g^{\mu\nu}} = \frac{a^2}{\sqrt{-\tilde{g}}} \frac{\delta F}{\,\delta \tilde{g}^{\mu\nu}} = a^2 \tilde{A}_{\mu\nu} \eqend{,}
\end{equation}
and we can equivalently show that $\tilde{A}_{\mu\nu}$ is of order $\kappa^2$. Another consequence of the invariance of $F$ under conformal transformations of $\tilde{g}_{\mu\nu}$ is the vanishing trace of $\tilde{A}_{\mu\nu}$:
\begin{equation}
0 = \frac{\delta F}{\delta a} = \frac{\delta F}{\delta \tilde{g}^{\mu\nu}} \frac{\partial \tilde{g}^{\mu\nu}}{\partial a} = - 2 \sqrt{-\tilde{g}} \tilde{A}_{\mu\nu} a^{-1} \tilde{g}^{\mu\nu} \eqend{,}
\end{equation}
so that $\tilde{g}^{\mu\nu} \tilde{A}_{\mu\nu} = 0$.

Expressing now the Weyl tensor in terms of the Riemann tensor and its contractions according to its defining equation~\eqref{weyl_tensor_definition}, we can write
\begin{equation}
\label{F_integral}
F = 2 \int \left[ \tilde{R}^{\mu\nu} \tilde{R}_{\mu\nu} - \frac{1}{3} \tilde{R}^2 \right] 
\sqrt{-\tilde{g}} \total^4 x
+ \int \mathcal{E}_4 \sqrt{-\tilde{g}} \total^4 x  \eqend{,}
\end{equation}
where $\mathcal{E}_4$ is the integrand of the four-dimensional Euler invariant and is given by
\begin{equation}
\label{E_integrand}
\mathcal{E}_4 = \tilde{R}^{\mu\nu\rho\sigma} \tilde{R}_{\mu\nu\rho\sigma}
- 4 \tilde{R}^{\mu\nu} \tilde{R}_{\mu\nu} + \tilde{R}^2
\eqend{.}
\end{equation}
The generalized Gauß-Bonnet theorem establishes that the integral of $\mathcal{E}_4$ is a topological invariant, namely $32\pi^2$ times the Euler characteristic. From eq.~\eqref{F_integral} and the fact that the variational derivative of a topological invariant vanishes, we see that $\tilde{A}_{\mu\nu}$ can be expressed entirely in terms of the Ricci tensor, the Ricci scalar and covariant derivatives acting on them.

We are now ready to use order reduction. Taking into account that
\begin{equation}
\label{rmunu_orderreduction}
\begin{split}
\tilde{R}_{\mu\nu} - \frac{1}{2} \tilde{g}_{\mu\nu} \tilde{R} + \Lambda \tilde{g}_{\mu\nu} &= \bigo{\kappa^2} \\
\tilde{R} - 4 \Lambda &= \bigo{\kappa^2} \\
\end{split}
\end{equation}
and substituting $\tilde{R}_{\mu\nu} = \Lambda \tilde{g}_{\mu\nu}$ into the expression for $\tilde{A}_{\mu\nu}$, the result can only be proportional to $\tilde{g}_{\mu\nu}$ up to order $\kappa^2$. However, since $\tilde{g}^{\mu\nu} \tilde{A}_{\mu\nu} = 0$, we conclude that $\tilde{A}_{\mu\nu}$ is of order $\kappa^2$ when order reduction is employed. One can alternatively check this fact by substituting eq.~\eqref{rmunu_orderreduction} into the explicit expression for $\tilde{A}_{\mu\nu}$ in terms of the Ricci tensor in eq.~\eqref{tensors_ab_definition}.

In conclusion, we see that we only have to consider the local terms in eq.~\eqref{einstein_semiclassical_conformal} when using order reduction.

%% file: nonperturbative_scalarvector.tex
The solutions of eq.~\eqref{semiclassical_perturbation_or_vector} for the components of the vector perturbation $v^\text{T}_i$ are arbitrary functions of time, which can be eliminated by a gauge transformation. Indeed, by using the residual gauge freedom described at the end of sec.~\ref{linear_gauge} and choosing $\xi^\text{T}_i$ as an appropriate function of time only, we can set $v^\text{T}_i = 0$.

For $\phi$, the solution of eq.~\eqref{semiclassical_perturbation_or_phi} is also an arbitrary function of time. The solution for $\psi$ is then given by
\begin{equation}
\psi = f(t) - \frac{1}{4 \eta} \left( 1 + \frac{4}{9} \nu \right) \phi(t) \, \vec{r}^2 \eqend{.}
\end{equation}
Since $\psi$ enters into the perturbation $h_{\mu\nu}$ only through a spatial derivative according to eq.~\eqref{perturbation_temporal_decompose}, the arbitrary function $f(t)$ does not change the perturbation $h_{\mu\nu}$ and we can set it to zero. If we want to start with bounded initial perturbations, we must exclude solutions which are unbounded and have to take $\phi = 0$.

On the other hand, if we had not excluded such unbounded solutions, we would first have to choose $\xi_0$ appropriately to make $\phi$ vanish, and would need to take a similar unbounded function $\xi = - 1/(2 \eta) \xi_0 \, \vec{r}^2$ so that the combination $\partial_i \partial_j \sigma + \tau \delta_{ij}$ which enters into the decomposition of the perturbation \eqref{perturbation_spatial_decompose} still vanishes. The solution for $\psi$ is then an arbitrary function of time which we can set to zero as above.

Thus, we see that when order reduction is employed, both vector and scalar parts after solving the constraints are pure gauge and can be eliminated by a residual gauge transformation of the kind mentioned at the end of sec.~\ref{linear_gauge}.

Note that Anderson et al.~\cite{anderson09}, who investigated scalar perturbations without using order reduction, also concluded that those perturbations (which they refer to as perturbations of the first kind) have to vanish because the corresponding solutions that they found lie outside the range of validity of the semiclassical theory. The gauge-invariant variables that they introduce, however, take a simple form in a gauge rather different from the one we employ, and so their intermediate expressions are not directly comparable to ours.

%% file: nonperturbative_tensor.tex
To find the solutions for the tensor perturbations, we take the Fourier transform with respect to the spatial coordinates,
\begin{equation}
\label{tensor_polariz}
h^\text{TT}_{ij}(\eta, \vec{x}) = \sum_{s=\pm} \int e_{ij}^s(\vec{p}) g_s(\eta, \vec{p}) \mathe^{\mathi \vec{p} \vec{x}} \frac{\total^3 p}{(2\mathpi)^3} \eqend{,}
\end{equation}
where $e_{ij}^\pm(\vec{p})$ are a pair of transverse and traceless tensors corresponding to two different polarizations. Equation~\eqref{semiclassical_perturbation_or_tensor} then becomes
\begin{equation}
\label{tensor_mode_eq}
g''_\pm - \frac{2}{\eta} (1-\nu) g'_\pm + (1-2\nu) \vec{p}^2 g_\pm = \bigo{\kappa^4} \eqend{.}
\end{equation}
Setting $\omega^2 = (1 - 2 \nu) \vec{p}^2$, $s = - \omega \eta$ and $g_\pm = s^{\frac{3}{2}-\nu} f_\pm(s)$, this reduces to a Bessel equation for $f_\pm$, whose general solution is
\begin{equation}
\label{tensor_mode_solution}
g_\pm = \left( - \omega \eta \right)^{\frac{3}{2}-\nu} \left[ C_1^\pm J_{\frac{3}{2}-\nu}(- \omega \eta) + C_2^\pm Y_{\frac{3}{2}-\nu}(- \omega \eta) \right] \eqend{,}
\end{equation}
where $C_1^\pm$ and $C_2^\pm$ are integration constants.

For the particular case $\vec{p}=0$, which corresponds to no spatial dependence in position space, eq.~\eqref{semiclassical_perturbation_or} can be solved directly and the general solution is given by
\begin{equation}
\label{tensor_mode_solution_p0}
h^\text{TT}_{ij} = D_{ij} (-\eta)^{3-2\nu} + E_{ij} \eqend{,}
\end{equation}
where $D_{ij}$ and $E_{ij}$ are traceless tensors with respect to the induced background metric of the spatially flat sections, and independent of the spatial coordinates and the conformal time.\footnote{These solutions can also be obtained by taking the limit $\omega \to 0$ of eq.~\eqref{tensor_mode_solution} after rewriting $C_1=\bar{C}_1/\omega^{3-2\nu}$ so that a finite non-vanishing limit is obtained for the solutions associated with the two integration constants. In addition, the different ways of taking the limit $\vec{p} \to 0$ of $e_{ij}^\pm(\vec{p})$ and the possibility of considering arbitrary linear combinations gives rise to the general traceless tensors $D_{ij}$ and $E_{ij}$.}
The first term on the right-hand side of eq.~\eqref{tensor_mode_solution_p0} corresponds to a Bianchi I anisotropic deformation of de Sitter, whereas the second one is pure gauge and can be eliminated by the residual gauge transformation generated by the transverse vector $\xi^\text{T}_i = E_{ij} x^j / 2$.

%% file: stability_riemann.tex
In this section we analyze the stability of the semiclassical de Sitter geometry with respect to small metric perturbations including the back-reaction due to quantum vacuum effects from conformal fields. We do so by focusing on the evolution of the Riemann tensor associated with the linearly perturbed metric $\tilde{g}_{\mu\nu}$, which has a number of appealing properties.
First of all, the linear perturbation around de Sitter of the Riemann tensor $\tilde{R}^{\alpha\beta}{}_{\gamma\delta}$ with appropriately raised indices is a gauge-invariant object. This follows from the fact that for the unperturbed background it can be written as $\tilde{R}^{(0)}{}^{\alpha\beta}{}_{\gamma\delta} = 2 H^2 \delta^{[\alpha}_{[\gamma} \delta^{\beta]}_{\delta]}$, whose Lie derivative with respect to an arbitrary vector field vanishes. 
Furthermore, with this index structure the components remain unchanged when rescaling by the same constant the basis vectors of the tangent space at a given point.
This implies that the components coincide with those in the physical basis of orthonormal vectors $\{ a^{-1} \partial_0, a^{-1} \partial_i \}$ of the background metric.
Finally, the Riemann tensor provides a suitable characterization of the local geometry, in terms of which the stability and attractor nature of semiclassical de Sitter spacetime can be naturally formulated, as further discussed at the end of this subsection. 

In terms of the metric perturbations the linearized Riemann tensor is given by
\begin{equation}
\label{riemann1}
\begin{split}
\tilde{R}^{(1)}{}^{\alpha\beta}{}_{\gamma\delta} &= 2 H^2 \delta^\alpha_{[\gamma} \delta_{\delta]}^\beta h_{00} + 2 H^2 \eta^2 \eta^{\mu[\alpha} \eta^{\beta]\nu} \partial_\nu \partial_{[\gamma} h_{\delta]\mu} \\
&\quad+ 2 H^2 \eta \delta^\mu_{[\gamma} \delta_{\delta]}^{[\beta} \eta^{\alpha]\nu} \left( 2 \partial_{(\mu} h_{\nu)0} - h'_{\mu\nu} \right) \eqend{.}
\end{split}
\end{equation}
Using the gauge transformation \eqref{gauge_trafo}, one can explicitly check that it is indeed gauge invariant at linear order.
Fourier transforming with respect to the spatial coordinates and specializing to tensor perturbations, we get
\begin{equation}
\label{riemann_in_perturbation}
\tilde{R}^{(1)}{}^{\alpha\beta}{}_{\gamma\delta} = 2 H^2 \int \left( {S^{[\alpha\beta]}_+}{}_{[\gamma\delta]} + {S^{[\alpha\beta]}_-}{}_{[\gamma\delta]} \right) \mathe^{\mathi \vec{p} \vec{x}} \frac{\total^3 p}{(2\mathpi)^3} \eqend{,}
\end{equation}
where
\begin{equation}
\label{riemann_decay}
\begin{split}
{S^{0j}_\pm}{}_{0k} &= - (e^\pm)^j_k \eta g'_\pm \\
{S^{ij}_\pm}{}_{0k} &= \mathi \eta p^i S^{0j}_\pm{}_{0k} \\
{S^{ij}_\pm}{}_{kl} &= p^i p_k (e^\pm)^j_l \eta^2 g_\pm + \delta^i_k S^{0j}_\pm{}_{0l} \eqend{.}
\end{split}
\end{equation}
Hence, we can see that all the Riemann components can be written in terms of $g_\pm$ and $g'_\pm$. Since everything that will be said is entirely equivalent for both polarizations, for ease of notation we will omit in the remainder of this section the subindices $\pm$ labeling the two transverse polarizations associated with each momentum $\vec{p}$.

Let us consider first the evolution of the Riemann tensor for modes well \emph{outside the horizon}, i.e.\ with $\abs{\omega\eta} \ll 1$. In this case one needs to evaluate the Bessel functions in eq.~\eqref{tensor_mode_solution} using eqs.~\eqref{small_x1} and \eqref{small_x2}, which leads to
\begin{equation}
\label{early_time}
\begin{split}
g &\sim - \frac{C_2}{\mathpi} \Gamma\left(\frac{3}{2}-\nu\right) 2^{\frac{3}{2}-\nu} + \bigo{\omega\eta} = \const + \bigo{\omega\eta} \eqend{,} \\
g' &\sim \omega \frac{C_2}{\mathpi} \Gamma\left( \frac{1}{2}-\nu \right) 2^{\frac{1}{2}-\nu} + \bigo{\omega\eta} = \const + \bigo{\omega\eta} \eqend{.}
\end{split}
\end{equation}
Substituting into eq.~\eqref{riemann_decay} we see that the components of the Riemann perturbation in a physical basis decay like $1/a = -H\eta$ or higher order at late times, i.e.\ in the limit $\eta \to 0$.

On the other hand, for modes \emph{inside the horizon}, with $\abs{\omega\eta} \gg 1$, one can use eqs.~\eqref{large_x1} and \eqref{large_x2} to see that $g$ and $g'$ are of the form
\begin{equation}
\begin{split}
g &\sim \sqrt{\frac{2}{\pi}} (-\omega\eta)^{1-\nu} \left[ 1 + \bigo{1/\omega\eta} \right] \eqend{,} \\
g' &\sim \omega \sqrt{\frac{2}{\pi}} (-\omega\eta)^{1-\nu} \left[ 1 + \bigo{1/\omega\eta} \right] \eqend{.}
\end{split}
\end{equation}
times an oscillatory factor corresponding to a linear combination of $\sin(\omega\eta)$ and $\cos(\omega\eta)$. Thus, from eq.~\eqref{riemann_decay} it follows that inside the horizon the components of the Riemann perturbation oscillate with an amplitude that decays like $1/a^{1-\nu}$.

Putting these results together we can conclude that de Sitter spacetime remains stable with respect to small metric perturbations of the semiclassical mean geometry when the quantum back-reaction of conformal matter fields is included. This is guaranteed by the fact that for any Fourier mode with comoving momentum $\vec{p}$ the perturbation of the Riemann tensor decays like $1/a^{1-\nu}$  (times an oscillatory factor) when the corresponding physical wavelength $2\pi a/\abs{\vec{p}}$ is smaller than the de Sitter radius $1/H$ and like $1/a$ when it is larger, together with the regularity of the perturbation around horizon crossing (when the wavelength is comparable to $1/H$). This extends the conclusions of the no-hair theorem for de Sitter spacetime, which is not only stable with respect to small metric perturbations but also a late-time attractor in classical general relativity, to the case where radiative corrections from loops of conformal fields are considered. In fact, the main effect of the radiative corrections compared to the classical case for pure gravity, which corresponds to taking $\nu=0$ in our results, is simply to alter slightly the exponent of the power-law decay for modes inside the horizon.

Our result can be used to illustrate in a simple way the fact that the stability and the character of late-time attractor of de Sitter spacetime applies to sufficiently localized observables characterizing the geometry within a region of fixed physical size (as opposed to comoving). The tensor perturbation $h^\text{TT}_{ij}(\eta, \vec{x})$ and the amplitude $g(\eta, \vec{p})$ associated with a given momentum and polarization are gauge-invariant objects with well-defined geometrical meaning. However, as their characteristic physical wavelength gets exponentially redshifted, at late times one would need to measure them over regions with a physical size that becomes arbitrarily large. Instead, the deviations of the geometric properties within a region of fixed physical size compared to those of de Sitter decay exponentially with the (proper) cosmological time.
These features are adequately captured by the behavior of the Riemann tensor, which provides a good characterization of the local geometry.

%% file: stability_perturbative.tex
In this subsection we will compare the exact nonperturbative solutions of the linearized semiclassical equation~\eqref{tensor_mode_eq}, given by eq.~\eqref{tensor_mode_solution}, to those that result from solving the equation perturbatively in $\nu$. We will see that the perturbative solutions cease to be valid when the modes evolve inside the horizon for a sufficiently long time.

Here we concentrate on initial perturbations corresponding to Bunch-Davies positive frequency modes, but our conclusions can be easily generalized to arbitrary initial conditions. This choice amounts to setting $C_2 = \mathi C_1$, so that the linear combination within the square bracket on the right-hand side of eq.~\eqref{tensor_mode_solution} becomes a Hankel function of the first kind, with a purely positive frequency oscillatory behavior at early times. Indeed, using eqs.~\eqref{large_x1} and \eqref{large_x2} one can see that in this case the behavior for $\abs{\omega\eta} \gg 1$ of the non-perturbative solution is given by
\begin{equation}
\label{nonperturbative_early_time}
g = C_1 \sqrt{\frac{2}{\mathpi}} \mathe^{- \mathi \frac{\mathpi}{2} (2-\nu)} \left( - \omega \eta \right)^{1-\nu} \mathe^{- \mathi \omega \eta} \left[ 1 + \bigo{1/\omega\eta} \right] \eqend{.}
\end{equation}
On the other hand, if we first expand the solution \eqref{tensor_mode_solution} in powers of $\nu$ employing eqs.~\eqref{nu_expansion1}--\eqref{nu_expansion2}, we get
\begin{equation}
\label{perturbative_expansion}
\begin{split}
g_\text{p} &= - C_1 \sqrt{\frac{2}{\mathpi}} \mathi \bigg[ \left( 1 + \mathi \abs{\vec{p}} \eta \right)
\mathe^{- \mathi \abs{\vec{p}} \eta} - \nu \left( 2 + \vec{p}^2 \eta^2 \right) \mathe^{- \mathi \abs{\vec{p}} \eta} \\
&\quad+ \mathi \nu \left[ \sin \left( \abs{\vec{p}} \eta \right) - \abs{\vec{p}} \eta \cos \left( \abs{\vec{p}} \eta \right) \right] \left[ \ln \left( \vec{p}^2 \eta^2 \right) - \mathi \mathpi \right] \\
&\quad+ \nu \left( 1 - \mathi \abs{\vec{p}} \eta \right) \left[ \Ein\left( - 2 \mathi \abs{\vec{p}} \eta \right) + \gamma + \ln 2 \right] \mathe^{\mathi \abs{\vec{p}} \eta} \bigg] \\
&\quad+ \bigo{\nu^2} \eqend{.}
\end{split}
\end{equation}

At late times (for $\abs{\omega\eta} \ll 1$) the exact solution is well approximated by the perturbative solution \eqref{perturbative_expansion}: one recovers the asymptotic behavior in eq.~\eqref{early_time} but with the constant term and the coefficients of the higher-order ones given by their expansion in powers of $\nu$ truncated at linear order. That is, however, not the case at early times, with $\abs{\omega\eta} \gg 1$.
This can be seen by considering the early-time limit of the perturbative solution \eqref{perturbative_expansion}:
\begin{equation}
\label{perturbative_divergence}
\begin{split}
g_\text{p} &\sim - C_1 \sqrt{\frac{2}{\mathpi}} \left(-\abs{\vec{p}} \eta\right) \mathe^{- \mathi \abs{\vec{p}} \eta} \times \\
&\times \left[ 1 - \nu \left( 1 - \frac{\mathi \mathpi}{2} + \ln \left( - \abs{\vec{p}} \eta \right)  - \mathi \abs{\vec{p}} \eta \right) + \bigo{\nu^2} \right] \eqend{,}
\end{split}
\end{equation}
which coincides with the result that one obtains by expanding in powers of $\nu$ the asymptotic expression \eqref{nonperturbative_early_time} for the exact solution.
It is clear that the perturbative solution deviates significantly from the exact one when $\nu \abs{\vec{p}} \abs{\eta} \gtrsim 1$.
The implications can be more easily understood if we normalize the modes so that they have a fixed amplitude at the initial time $\eta_0$ independently of the particular value of $\eta_0$, which can be implemented by dividing the mode by its value at $\eta_0$.
Proceeding in this way with the exact solution, one finds for example that the amplitude of a mode which was initially well  within the horizon has decreased by a factor $1/(-\omega\eta_0)^{1-\nu}$ by the time of horizon crossing. In contrast, repeating the procedure with the perturbative solution and treating $\nu$ perturbatively when normalizing by the value at $\eta_0$, one obtains an amplitude at horizon crossing of order $(-1/\abs{\vec{p}} \eta_0) \left( 1 + \nu \ln(-\abs{\vec{p}} \eta_0) - \mathi \nu \abs{\vec{p}} \eta_0 \right)$.
For modes which have spent a long time inside the horizon, so that $\nu \abs{\vec{p}} \abs{\eta_0} \gtrsim 1$, this amplitude at horizon crossing can be significantly larger.

The reason for the potentially large deviation of the perturbative solution with respect to the exact one is a breakdown of perturbation theory: the actual condition for the validity of the perturbative solution is $\nu \abs{\vec{p}} \eta_0 \ll 1$, which can be violated even for $\nu \ll 1$ when considering $\abs{\vec{p}} \abs{\eta_0}$ large enough. The term proportional to $\mathi \nu \abs{\vec{p}} \eta$ in eq.~\eqref{perturbative_divergence}, responsible for the main deviations, is a secular term arising from the truncated perturbative expansion in powers of $\nu$ of the oscillatory factor $\mathe^{- \mathi \omega \eta}$ in eq.~\eqref{nonperturbative_early_time}. Such a breakdown of perturbation theory for long times is very common when determining the evolution of a system by solving perturbatively the corresponding dynamical equations.
This can be illustrated with the simple example of a harmonic oscillator with frequency $\Omega + \delta\Omega$. If one solves perturbatively in $\delta\Omega$ the corresponding  equation of motion, $\ddot{x} + (\Omega + \delta\Omega)^2 x = 0$, to first order one finds a solution of the form $x(t) \approx A \sin(\Omega t + \varphi) [ 1 + \delta\Omega t ]$, where the correction grows with time. In this case, however, the exact solution is obviously known: $x(t) = A \sin\big[(\Omega + \delta\Omega) t + \varphi \big]$. It is, therefore, clear that the exact solution is qualitatively very similar to the unperturbed one but with a slightly corrected frequency. The growing terms in the perturbative solution, which are commonly know as secular terms, reflect the fact that the perturbative solution is not valid for arbitrarily long times but restricted instead to times such that $\delta\Omega \, t \ll 1$.
This is a rather simple example, but the situation is completely analogous for our semiclassical solution.

We close this subsection by comparing the perturbative solution obtained by expanding the exact solution, which has been discussed above, with the one obtained by solving perturbatively eq.~\eqref{tensor_mode_eq} as done in ref.~\cite{fordetal}. Substituting the classical solution for $g$ into the terms proportional to $\nu$ in eq.~\eqref{tensor_mode_eq}, treating them as a source, and using the retarded propagator for the unperturbed equation (with $\nu=0$), one gets the following result for the perturbative correction, already obtained in ref.~\cite{fordetal} (see their eq.~(48)):
\begin{widetext}
\begin{equation}
\label{ford_correction1}
g^{(1)} \propto 2 \mathi \, \nu \int_{\eta_0}^\eta \left[ \left( 1 + \vec{p}^2 \eta \eta' \right) \sin \left[ \abs{\vec{p}} (\eta - \eta') \right] - \abs{\vec{p}} (\eta-\eta') \cos \left[ \abs{\vec{p}} (\eta - \eta') \right] \right] \frac{\mathe^{- \mathi \abs{\vec{p}} \eta'}}{\eta'} \total \eta' \eqend{.}
\end{equation}
\end{widetext}
The integral can be done exactly with the help of appendix~\ref{appendix_special}. For large negative values of $\eta_0$ one gets 
\begin{equation}
\label{ford_correction2}
g^{(1)} \sim \nu \left[ - \mathi \abs{\vec{p}} \eta_0 + \ln\left( - \abs{\vec{p}} \eta_0 \right)
+ \frac{1}{2} \mathe^{- 2 \mathi \abs{\vec{p}} \eta_0} + \bigo{1} \right] \eqend{,}
\end{equation}
where we have 
set $\eta=0$ for the upper limit, whose effect is unimportant for a late-time solution. (Note that the logarithmic term and the phase were not included in ref.~\cite{fordetal}, where only the dominant term was retained.) If we had normalized the classical solution at the initial time, which essentially amounts to dividing the unperturbed solution by $1/\eta_0$, the perturbative correction in eq.~\eqref{ford_correction2} would agree with the perturbative solution \eqref{perturbative_divergence} when normalized at the initial time $\eta_0$ as discussed below that equation. There is actually a slight discrepancy: one does not get the term $\mathe^{- 2 \mathi \abs{\vec{p}} \eta_0}/2$ appearing in eq.~\eqref{ford_correction2}. This is due to a different choice of initial conditions. Whereas our exact solution \eqref{nonperturbative_early_time} results from choosing $C_2=\mathi C_1$ exactly (at all orders in perturbation theory), eqs.~\eqref{ford_correction1}--\eqref{ford_correction2} correspond to making the same choice for the zeroth-order solution but requiring the first-order correction and its derivative to vanish at the initial time. By imposing the same conditions when determining our exact and perturbative solutions, one obtains a new pair of constants $C_1$ and $C_2$ which differ at order $\nu$ and lead to a perturbative solution in full agreement with eq.~\eqref{ford_correction2}.

It should be pointed out that the perturbative correction can be large only when the corresponding mode was trans-Planckian at the initial time, i.e.\ its physical wavelength was much smaller than the Planck length (otherwise one has $- \nu \abs{\vec{p}} \eta_0 \ll 1$ and the perturbative correction is always small), as recognized in ref.~\cite{fordetal}. At such scales semiclassical gravity is not guaranteed to provide an accurate description. Furthermore, one would need a rather small amplitude of the initial perturbations so that nonlinear gravitational effects do not become important: otherwise for such short wavelengths the effective stress tensor quadratic in the metric perturbations \cite{brill64,isaacson68} can generate a strong back-reaction on the background expansion, and even dominate over the cosmological constant.
In any case, even if one carries out a linearized analysis without much concern for these issues, as done in ref.~\cite{fordetal}, the corrections to the classical solution will always be small as shown by our exact solutions and discussed above.

%% file: initial_states.tex
So far we have considered the dynamics of metric perturbations, but for a fixed initial state of the matter field, namely the Bunch-Davies vacuum. 
The state-dependent expectation value of the stress tensor is affected by the metric perturbations, but this is due to the their effect on the evolution of the scalar field operator in the Heisenberg picture or, alternatively, on the evolution of the state in the Schrödinger picture.
However, if one wants to allow changes in the initial state, eq.~\eqref{einstein_semiclassical_conformal} needs to be generalized.
In fact, as discussed in appendix~\ref{regular_states}, the Bunch-Davies vacuum of the matter fields is no longer a Hadamard state (free of excitations at arbitrarily short wavelengths) for nonvanishing metric perturbations at the initial time: Therefore, the initial state needs to be modified so that it is a Hadamard state in that case and, in particular, the renormalized expectation value of the stress tensor is finite at the initial time.

Starting with eq.~\eqref{einstein_semiclassical} and considering not only linear perturbations around the de Sitter metric but also small perturbations of the initial state, one obtains
\begin{widetext}
\begin{equation}
\label{semiclassical_inhomogeneous}
\tilde{G}_{\mu\nu}^{(1)} + \Lambda a^2 h_{\mu\nu} - a_1 \tilde{A}_{\mu\nu}^{(1)} - a_2 \tilde{B}_{\mu\nu}^{(1)} - \frac{1}{2} \kappa^2 \expect{\op{T}_{\mu\nu}^{(1)}}_\text{ren} = \frac{1}{2} \kappa^2 \delta \expect{\op{T}_{\mu\nu}^{(0)}}_\text{ren} \equiv \frac{1}{2} \kappa^2 \delta T_{\mu\nu} \eqend{,}
\end{equation}
\end{widetext}
where the the right-hand side corresponds to the perturbation of the expectation value of the stress tensor evaluated on the background metric due to the perturbation of the initial state. Here it has been assumed that such a term is of the same order as the remaining terms, which are linear in the metric perturbations. Hence, no further terms should be considered since those corresponding to the perturbation of the initial state in the stress tensor expectation value evaluated on the perturbed metric would be of higher order.

Note that eq.~\eqref{semiclassical_inhomogeneous} and the procedure employed below can also be used to consider the effect of any other additional stress tensor sources (even classical ones) which can be treated perturbatively and regarded of the same order as the terms linear in the metric perturbations.

For conformal fields in a FLRW background eq.~\eqref{semiclassical_inhomogeneous} for the linear metric perturbations reduces again to eq.~\eqref{einstein_semiclassical_conformal} plus the source term involving $\delta T_{\mu\nu}$.
Furthermore, one can introduce a decomposition of $\delta T_{\mu\nu}$ analogous to that for the metric perturbations in eqs.~\eqref{perturbation_spatial_decompose}-\eqref{perturbation_temporal_decompose}, which is complemented now by the conservation requirement for $\delta T_{\mu\nu}$ with respect to the background metric, i.e.\ ${}^{(0)}\nabla^\mu \delta T_{\mu\nu} = 0$ with ${}^{(0)}\nabla_\mu$ being the covariant derivative associated with the background metric $g^{(0)}_{\mu\nu} = a^2 \eta_{\mu\nu}$.
More specifically, we can write
\begin{equation}
\label{deltaT}
\begin{split}
\delta T_{0i} &= \delta T_{0i}^\text{T} + \partial_i \chi \\
\delta T_{ij} &= \delta T^\text{TT}_{ij} + 2 \partial_{(i} W_{j)}^\text{T}
+ \left( \partial_i \partial_j - \frac{1}{3} \laplace \right) \rho + \frac{\delta T}{3} \delta_{ij}
\eqend{,}
\end{split}
\end{equation}
and take as independent quantities the transverse and traceless tensor $\delta T^\text{TT}_{ij}$, the transverse vector $\delta T_{0i}^\text{T}$ and the two scalar functions $\delta T_{00}$ and $\delta T$.
The temporal component of the conservation equation determines the scalar $\chi$ and, similarly, the scalar part of the spatial projection of the conservation equation, which can be written as the gradient of a function, determines the scalar $\rho$. On the other hand, the transverse part of this spatial projection determines the transverse vector $W^\text{T}_j$.
Applying this decomposition to the generalization of eq.~\eqref{einstein_semiclassical_conformal} including the source $\delta T_{\mu\nu}$, one obtains eqs.~\eqref{semiclassical_scalar}--\eqref{semiclassical_tensor} plus the corresponding sources. Finally, using order reduction eqs.~\eqref{semiclassical_perturbation_or} are generalized to
\begin{widetext}
\begin{subequations}
\begin{align}
h^{\prime\prime\text{TT}}_{ij} - \frac{2}{\eta} \left( 1 - \nu \right) h^{\prime\text{TT}}_{ij} - \left( 1 - 2 \nu \right) \laplace h^\text{TT}_{ij} &= \kappa^2 \delta T^\text{TT}_{ij} + \bigo{\kappa^4}
\label{in_state_eqs1} \eqend{,} \\
\laplace v^\text{T}_i &= - \kappa^2 \delta T^\text{T}_{0i} + \bigo{\kappa^4} \label{in_state_eqs2} \eqend{,} \\
\laplace \phi &= - \kappa^2 \left( \delta T_{00} - \frac{1}{2} \eta \delta T'_{00} + \frac{1}{2} \eta^{\mu\nu} \delta T_{\mu\nu} \right) + \bigo{\kappa^4} \label{in_state_eqs3} \eqend{,} \\
\laplace \psi + \frac{3}{2 \eta} \left( 1 + \frac{4}{9} \nu \right) \phi &= \frac{1}{4} \eta \kappa^2 \delta T_{00} + \bigo{\kappa^4} \label{in_state_eqs4} \eqend{,}
\end{align}
\end{subequations}
\end{widetext}
which correspond, respectively, to the transverse and traceless part of the $ij$ component of the order-reduced equations, the transverse part of the $0i$ component, the spatial divergence of the $0i$ component, and the $00$ component. Moreover, the conservation equation has been used to express the spatial divergence of $\delta T_{0i}$ in terms of $\delta T_{00}$ and $\delta T$ on the right-hand side of eq.~\eqref{in_state_eqs3}.

%% file: initial_tensor.tex
Let us start with eq.~\eqref{in_state_eqs1} for the tensor perturbations. In addition to employing eq.~\eqref{tensor_polariz} for $h^\text{TT}_{ij}$, one can use the analogous expression
\begin{equation}
\label{stress_tensor_polariz}
\begin{split}
\delta T^\text{TT}_{ij}(\eta, \vec{x}) = \sum_{s=\pm} \int e_{ij}^s(\vec{p}) J_s(\eta, \vec{p})
\mathe^{\mathi \vec{p} \vec{x}} \frac{\total^3 p}{(2\mathpi)^3}
\end{split}
\end{equation}
for the stress tensor perturbation and obtains the following equation for each one of the two polarizations:
\begin{equation}
\label{tensor_inhom_eq}
g''_\pm - \frac{2}{\eta} \left( 1 - \nu \right) g'_\pm + \left( 1 - 2 \nu \right) \vec{p}^2 g_\pm
= \kappa^2 J_\pm \eqend{.}
\end{equation}
This equation is identical to eq.~\eqref{tensor_mode_eq} but with an inhomogeneous source term.
The general solution can be written as $g_\pm = g_\text{h}^\pm + g_\text{i}^\pm$, a sum of a homogeneous solution containing the information on the initial conditions at the time $\eta_0$ and a particular solution of the inhomogeneous equation with vanishing initial conditions,
\begin{equation}
\label{tensor_inhom_sol1}
g_\text{i}^\pm (\eta) = \int_{\eta_0}^0 G_\text{ret}(\eta,\eta') J_\pm(\eta') \total\eta' \eqend{,}
\end{equation}
where $G_\text{ret}$ is the retarded propagator associated with eq.~\eqref{tensor_inhom_eq}. Given two independent homogeneous solutions, $u_1$ and $u_2$, the retarded propagator for such a linear second-order differential equation can be expressed as
\begin{equation}
\label{retarded_prop}
G_\mathrm{ret}(\eta,\eta') = \frac{u_1(\eta) u_2(\eta') - u_2(\eta) u_1(\eta')}{W(\eta')} \theta(\eta-\eta') \eqend{,}
\end{equation}
where $W(\eta') = u'_1(\eta') u_2(\eta') - u'_2(\eta') u_1(\eta')$ is the Wronskian for this pair of solutions, which is nonzero for independent solutions. Therefore, the inhomogeneous solution can be written as
\begin{equation}
\label{tensor_inhom_sol2}
g_\text{i}^\pm (\eta) = C_2^\pm(\eta) u_1(\eta) - C_1^\pm(\eta) u_2(\eta)
\end{equation}
with 
\begin{equation}
\label{tensor_inhom_sol3}
C_j^\pm (\eta) = \int_{\eta_0}^\eta \frac{u_j(\eta')}{W(\eta')} J_\pm(\eta') \total\eta' \eqend{.}
\end{equation}

Provided that $C_j^\pm (\eta)$ are regular and have a finite limit when $\eta \to 0$, as we will discuss next, the late-time behavior of the contribution form the inhomogeneous solution $g_\text{i}^\pm$  is the same as for the homogeneous solution, already analyzed in sec.~\ref{stability_riemann}.
We can consider the same pair of homogenous solutions as in that section and choose
\begin{equation}
\label{tensor_indep_sols}
\begin{split}
u_1(\eta) &= (- \omega \eta)^{\frac{3}{2}-\nu} J_{\frac{3}{2}-\nu}(- \omega \eta) \eqend{,} \\
u_2(\eta) &= (- \omega \eta)^{\frac{3}{2}-\nu} Y_{\frac{3}{2}-\nu}(- \omega \eta) \eqend{.}
\end{split}
\end{equation}
Taking into account that for this choice $W(\eta')=(2\omega/\pi) (-\omega\eta')^{2-2\nu}$, we can immediately see that $C_j^\pm (\eta)$ will be regular and have a finite limit when $\eta \to 0$ provided that $J_\pm(\eta')$ is also regular and decays faster than $(-\eta')^{1-2\nu}$ as that limit is approached.
This means that the components of the stress tensor $\delta T_{ij}^\text{TT}$ should decay slightly faster than $1/a$ in conformal coordinates, or $1/a^3$ in a physical basis. Such kind of behavior is fulfilled by classical radiation, which decays like $1/a^4$, and it seems plausible that small excitations of the vacuum state for the conformal fields considered here also exhibit the same decay; it is indeed the case for the class of regular states considered in appendix~\ref{regular_states}, as shown there.
We stress again that $\delta T_{\mu\nu}$ has been assumed to be small enough so that it is at most of the same order as the terms linear in the metric perturbation in eq.~\eqref{semiclassical_inhomogeneous} and nonlinear gravitational effects are not important (otherwise one could consider for instance an excitation with sufficiently large energy overdensity to form a black hole by gravitational collapse, contradicting the conclusions above).

As seen above, if the quantities in eq.~\eqref{tensor_inhom_sol3} have a regular limit as $\eta \to 0$, the inhomogeneous solutions $ g_\text{i}^\pm$ behave in the same way as the homogeneous ones and the same conclusions drawn in sec.~\ref{stability_secular} apply to the inhomogeneous solution as well.

%% file: initial_scalar_vector.tex
Unlike tensor perturbations, vector and scalar ones do not give any nontrivial contribution in the classical case, but they get a nonvanishing semiclassical correction.
Nevertheless, since they are governed by dynamical constraints which become simple algebraic equations when working in Fourier space for the spatial coordinates, their long time behavior can be directly determined from the fall-off properties of the stress tensor perturbation $\delta T_{\mu\nu}$.

Let us consider the \emph{vector} perturbations first. From eq.~\eqref{in_state_eqs2} it follows that in spatial Fourier space $\tilde{v}^\text{T}_i = (\kappa^2 / \vec{p}^2) \, \delta \tilde{T}^\text{T}_{0i}$ and the metric perturbations decay in the same way as the stress tensor.
From eq.~\eqref{riemann1} it follows that their contribution to the linear perturbation of the Riemann tensor is given by
\begin{equation}
\begin{split}
\tilde{R}^{(1)}{}^{0j}{}_{0l} &= -\frac{\mathi}{2} H^2 \eta \left( \eta p^j v^{\prime\text{T}}_l + \eta p_l v^{j\prime\text{T}} - p_l v^{j\text{T}} - p^j v^\text{T}_l \right) \eqend{,} \\
\tilde{R}^{(1)}{}^{0j}{}_{kl} &= H^2 \eta^2 p^j p_{[k} v^\text{T}_{l]} \eqend{,} \\
\tilde{R}^{(1)}{}^{ij}{}_{0l} &= - H^2 \eta^2 p^{[i} p_l v^{j]\text{T}} \eqend{,} \\
\tilde{R}^{(1)}{}^{ij}{}_{kl} &= 2 \mathi H^2 \eta \delta^{[j}_{[l} \left( p_{k]} v^{i]\text{T}} + p^{i]} v^\text{T}_{k]} \right) \eqend{,} \\
\end{split}
\end{equation}
and we can conclude that at late-times the curvature perturbations decay like $1/a \, \delta T^\text{T}_{0i}$ and, hence, for any decaying (or even asymptotically constant) stress tensor perturbation the stability and attractor character of de Sitter spacetime are not altered.

The situation is similar for \emph{scalar} perturbations but with some slight differences. From eqs.~\eqref{in_state_eqs3}--\eqref{in_state_eqs4} one can see that $\phi$ decays at least like $\delta T_{00}$ or $\eta^{\mu\nu} \delta T_{\mu\nu}$, whereas $\psi$ is only guaranteed to behave at late times like $a\, \delta T_{00}$ or $a\, \eta^{\mu\nu} \delta T_{\mu\nu}$.
Given the contributions of scalar perturbations to the linearized Riemann tensor,
\begin{equation}
\begin{split}
\tilde{R}^{(1)}{}^{0j}{}_{0l} &= \frac{1}{2} H^2 \left( 2 \delta^j_l \phi - \delta^j_l \eta \phi'
- \eta^2 p^j p_l \phi \right) \\
&\quad- H^2 \eta p^j p_l \left( \psi - \eta \psi' \right) \eqend{,} \\
\tilde{R}^{(1)}{}^{0j}{}_{kl} &= \mathi H^2 \eta \delta^j_{[k} p_{l]} \phi \eqend{,} \\
\tilde{R}^{(1)}{}^{ij}{}_{0l} &= - \mathi H^2 \eta \delta_l^{[i} p^{j]} \phi \eqend{,} \\
\tilde{R}^{(1)}{}^{ij}{}_{kl} &= 2 H^2 \delta^{[i}_{[k} \left( \delta_{l]}^{j]} \phi + 2 \eta p_{l]} p^{j]} \psi \right) \eqend{,}
\end{split}
\end{equation}
one can see that they basically fall off like $\delta T_{\mu\nu}$ at late times. Therefore, if $\delta T_{\mu\nu}$ decays at least like $1/a$ the conclusions about the stability of de Sitter space remain unchanged. Moreover, when it falls off as $1/a^2$, like classical radiation or as shown to be the case for the class of initial state perturbations described in appendix~\ref{regular_states}, the curvature perturbations due to scalar and vector perturbations fall off faster (by a factor $1/a$) than those due to tensor perturbations, which become the dominant contribution at late times.

%% file: general.tex
The results found for free conformal scalar fields in the previous sections can be straightforwardly generalized to any CFT for the matter sector (even strongly coupled ones). This is because when the background $g^{(0)}_{\mu\nu}$ is Minkowski spacetime, the key ingredient for obtaining the linearized stress tensor expectation value $\expect{\op{T}^{(1)}_{\mu\nu}[g^{(0)}+h]}_\text{ren}$, and hence the right-hand side of eq.~\eqref{einstein_semiclassical_conformal}, is entirely determined (up to a constant factor) by conformal as well as Poincaré invariance. Moreover, it transforms in a relatively simple way under conformal transformations and can be easily extended to the case of metric perturbations around a FLRW background, as described in some more detail below.

The stress tensor expectation value can be obtained by functionally differentiating the so-called CTP effective action $S_\text{eff}[g,g']$ \cite{martin99,mottola86,roura08}:
\begin{equation}
\label{stress-tensor}
\expect{\op{T}_{\mu\nu}[g]}_\text{ren} = \frac{2}{\sqrt{-g}} \left. \frac{\delta}{\delta g^{\mu\nu}} S_\text{eff}[g,g'] \right|_{g'=g} \eqend{.}
\end{equation}
The effective action $S_\text{eff}[g,g']$ can be written as
\begin{equation}
\label{CTPEA}
S_\text{eff}[g,g'] = S_\text{div}[g] - S_\text{div}[g'] + \Sigma[g,g'] \eqend{,}
\end{equation}
where the third term which includes nonlocal contributions results from functionally integrating out the matter fields. Performing those path integrals for the matter fields gives rise to UV divergences and one needs to introduce an appropriate regularization procedure; dimensional regularization is a good choice because it is compatible with general covariance (and in a number of cases with conformal invariance as well). Such divergences in $\Sigma[g,g']$ can be absorbed by local counterterms in the bare gravitational action. For massless fields the cosmological constant and the Einstein-Hilbert term do not get renormalized in dimensional regularization and only counterterms quadratic in the curvature are necessary, which have been denoted by $S_\text{div}[g]$ in eq.~\eqref{CTPEA}.
More specifically, the gravitational counterterms for a generic CFT on a curved background \cite{birrelldavies,mazur01} are given in dimensional regularization by
\begin{equation}
\label{counterterms_CFT}
\begin{split}
S_\text{div}[g] = \frac{1}{d-4} \bigg[ & b \int C_{\mu\nu\rho\sigma} C^{\mu\nu\rho\sigma} \sqrt{-g} \total^d x \\
&+ b' \int \mathcal{E}_4 \sqrt{-g} \total^d x \bigg] \eqend{,}
\end{split}
\end{equation}
where $d$ denotes the spacetime dimension and $\mathcal{E}_4$ is the integrand of the Euler invariant defined in eq.~\eqref{E_integrand}.
Although the (regulated) bare $\Sigma[g,g']$ is invariant under conformal transformations of the metric%
\footnote{Theories whose classical action is conformally invariant only in four dimensions can give rise to a finite counterterm quadratic in the Ricci scalar, which simply implies a finite change of the arbitrary parameter $\beta$ in eq.~\eqref{einstein_semiclassical_conformal}.} (see the appendix in ref.~\cite{eftekharzadeh12}),
the counterterms in $S_\text{div}[g]$ are not, which is the origin of the trace anomaly. In fact, the constant parameters $b$ and $b'$ in eq.~\eqref{counterterms_CFT}, which take specific values for each CFT, are also the coefficients of the Weyl-squared and $\mathcal{E}_4$ terms in the trace anomaly (the nonvanishing trace of the quantum stress tensor for conformal fields).
So far the statements in this paragraph are valid for an arbitrary metric $g$. If one is, however, interested in the linearized stress tensor for metric perturbations around a background $g^{(0)}_{\mu\nu}$ one only needs to consider terms in $S_\text{eff}[g,g']$ quadratic in the metric perturbations or, equivalently, focus on its second functional derivative,
which is related to the two-point function of the stress tensor in the background geometry.

Let us now show the universality of the linearized stress tensor expectation value in two steps.
The first step is to establish it for metric perturbations around a Minkowski background. By requiring conservation of the linearized stress tensor together with Poincaré and conformal invariance, it has been shown \cite{osborn94} that the form of the properly renormalized two-point function of the stress tensor in a Minkowski background for any conformal theory is essentially unique (up to a constant factor), and so is the renormalized vacuum expectation value of the stress tensor in the linearly perturbed metric.
Note that for perturbations around a Minkowski background the second term on the right-hand side of eq.~\eqref{counterterms_CFT} is at least cubic in the metric perturbations and does not contribute to the renormalized expectation value of the linearized stress tensor operator: it, therefore, depends only on the constant $b$.

The second step is to extend this result to the case of perturbations around a FLRW background via a conformal transformation. As already mentioned above, the counterterms in eq.~\eqref{counterterms_CFT} are not invariant under a conformal transformation of the metric and lead to the following change of the effective action:
\begin{equation}
\label{counterterms_diff}
\begin{split}
S_\text{eff}[\tilde{g},\tilde{g}'] &= S_\text{eff}[g,g'] - \left( S_\text{div}[g] - S_\text{div}[g'] \right) \\
&\quad+ \left( S_\text{div}[\tilde{g}] - S_\text{div}[\tilde{g}'] \right) \eqend{.}
\end{split}
\end{equation}
A conformal transformation leaves the two terms within the square brackets on the right-hand side of eq.~\eqref{counterterms_CFT} invariant in four dimensions and gives rise to terms of order $(d-4)$ otherwise, which amounts to a finite contribution in eq.~\eqref{counterterms_CFT}. The difference between the gravitational counterterms in two conformally related geometries is, therefore, finite. The difference between the Weyl-squared terms gives a term proportional to $C_{\mu\nu\rho\sigma} C^{\mu\nu\rho\sigma} \ln a$ in the effective action, whose functional derivative corresponds to the first term inside the second square bracket on the right-hand side of eq.~\eqref{einstein_semiclassical_conformal}. On the other hand, since the $\mathcal{E}_4$ term in eq.~\eqref{counterterms_CFT} vanishes up to quadratic order in the metric perturbations for perturbations around Minkowski space, only the $\tilde{\mathcal{E}}_4$ term for the conformally transformed metric $\tilde{g}$ contributes at this order. Its functional derivative corresponds to the terms inside the first square bracket on the right-hand side of eq.~\eqref{einstein_semiclassical_conformal}, which are conserved because they result from functionally differentiating a diffeomorphism-invariant integral.
In addition to a finite contribution to the coefficient of the Weyl-squared term, which can be absorbed by redefining the renormalization scale $\bar{\mu}$, the coefficient of the squared Ricci scalar term can take arbitrary finite values. Its functional derivative corresponds to the first term on the right-hand side of eq.~\eqref{einstein_semiclassical_conformal} and it generates a term $\tilde{\dalembert} \tilde{R}$ in the trace anomaly.

In summary, the linearized semiclassical equation~\eqref{einstein_semiclassical_conformal} will have the same form for any CFT and only the numerical coefficients in front of the two square brackets will change depending on the values of the parameters $b'$ and $b$, respectively.

%% file: conclusions.tex
Employing the method of order reduction, we have solved nonperturbatively the semiclassical Einstein equation governing the dynamics of linear metric perturbations around de Sitter spacetime when the quantum back-reaction of conformal scalar fields on the mean geometry is included.
Our exact solutions establish the stability of de Sitter with respect to general linear metric perturbations (of scalar, vector and tensor type) and extend some of the existing ``no-hair'' results for de Sitter in classical general relativity to the case in which the effects of the quantum vacuum polarization of conformal fields on the semiclassical geometry are included. Indeed, we confirm the late-time attractor character of de Sitter space (for geometrical properties within a region of fixed physical size) by showing that the perturbations of the Riemann tensor, which characterizes entirely the local geometry, fall off with an inverse power of the scale factor.

Perturbative solutions of the semiclassical equation for tensor perturbations have recently been obtained \cite{fordetal} and it was found that the correction to the classical solution can grow arbitrarily large for modes spending a long time inside the horizon. In contrast, our exact nonperturbative solutions exhibit oscillations with decaying amplitude inside the horizon and reveal a breakdown of perturbation theory for long times inside the horizon due to secular terms that arise when expanding perturbatively oscillatory factors with a perturbatively 
corrected frequency.
In addition, we have considered the effects due to perturbations of the initial state of the matter fields.
In fact, in order for the state of the fields to continue being a Hadamard state (with no unphysical excitations of arbitrarily short-wavelength modes) when metric perturbations are present at the initial time, it is in general necessary to correct the states that one would consider in the absence of perturbations and consider instead properly ``dressed'' states which are adiabatic on the perturbed geometry at sufficiently high order.

As explained in sec.~\ref{general}, our results are applicable to general conformal field theories with arbitrarily strong self-interaction couplings.
It would also be interesting to extend the results for free scalar fields to nonvanishing masses and arbitrary curvature couplings, particularly to massless (or sufficiently light) minimally coupled fields, which typically give rise to larger IR effects in de Sitter. 
We plan to return to this question in future investigations.

Besides semiclassical perturbations of the mean geometry, it is interesting to study the quantum fluctuations around it, which can be achieved by quantizing the metric perturbations around the mean geometry and dealing with them as an effective field theory. This was done in ref.~\cite{frv2011a}, where the two-point quantum correlation function for tensor metric perturbations around de Sitter, including one-loop corrections from conformal fields, was calculated. A suitable method for selecting the adiabatic vacuum of the interacting theory was employed and two-point functions compatible with de Sitter invariance were obtained. In particular, all secular terms cancelled out despite being a perturbative calculation. The reason for that can be understood with a simpler but qualitatively similar example.
Let us consider a system whose dynamics is governed by a  time-independent Hamiltonian with a small interaction term, and let us focus on the following two-point correlation function:
\begin{equation}
\label{example_correlation}
\begin{split}
C(t_2,t_1) &= \bra{\Psi_0} \hat{U}^\dagger(t_1,t_0) \op{U}^\dagger(t_2,t_1) \times \\
&\qquad \times B(t_2) \op{U}(t_2,t_1) A(t_1) \op{U}(t_1,t_0) \ket{\Psi_0} \eqend{,}
\end{split}
\end{equation}
where $A$ and $B$ are time-local operators in the Schrödinger picture. Evolving an arbitrary initial state $\ket{\Psi_0}$ for a long period $(t_1-t_0)$ will require in general a nonperturbative calculation. However, for eigenstates of the full Hamiltonian (including the small interaction term) one simply gets a phase factor, $\hat{U}(t_1,t_0) \ket{\Psi_0} = \mathe^{-\mathi E (t_1-t_0)} \ket{\Psi_0}$, which cancels exactly with the complex conjugate counterpart arising from the evolution of $\bra{\Psi_0}$.
Such an exact cancellation also implies a cancellation at every order in perturbation theory and guarantees the cancellation of any possible secular terms associated with the period $(t_1-t_0)$ which may arise in a perturbative calculation at finite order provided that an energy eigenstate of the full Hamiltonian is properly selected as the initial state.
Nevertheless, additional secular terms may arise due to the time evolution $\op{U}^\dagger(t_2,t_1)$ in case of large time differences between the arguments of the correlation function. Thus, a perturbative calculation of the ground-state correlation function will be valid for small $(t_2-t_1)$ no matter how large $(t_1-t_0)$ is, but will break down for large $(t_2-t_1)$.
The situation is analogous for metric perturbations around de Sitter.
In general a nonperturbative calculation is needed to evolve arbitrary initial states other than the adiabatic vacuum for a sufficiently long time, but it is also required in order to calculate loop corrections to the two-point function with respect to the adiabatic vacuum which are valid for large invariant intervals (both for time-like or space-like separations).
In this respect, the connection provided by stochastic gravity \cite{stochasticgravity} between the solutions of the semiclassical Einstein equation and the two-point quantum correlation functions for metric perturbations with (resummed) matter loop corrections \cite{hu04} suggests that the nonperturbative semiclassical solutions found here, and the methods employed to obtain them, could be exploited to compute matter loop corrections to the two-point function of the metric perturbations valid for arbitrarily large separations.

%% file: appendix_metric.tex
Given the perturbed metric $g_{\mu\nu}$ and expanding through quadratic order in the metric perturbation we have
\begin{equation}
\begin{split}
g_{\mu\nu} &= \eta_{\mu\nu} + h_{\mu\nu} , \\
g^{\mu\nu} &= \eta^{\mu\nu} - h^{\mu\nu} + h^\mu_\sigma h^{\nu\sigma} + \bigo{h^3} , \\
h &= \eta^{ab} h_{ab} , \\
\sqrt{-g} &= 1 + \frac{1}{2} h + \frac{1}{8} h^2 - \frac{1}{4} h_{\mu\nu} h^{\mu\nu} + \bigo{h^3} \eqend{,}
\end{split}
\end{equation}
where indices are raised and lowered with the unperturbed metric $\eta_{\mu\nu}$, i.e.\ we regard $h_{\mu\nu}$ as a tensor field in flat space. For the Christoffel symbols we get
\begin{equation}
\begin{split}
\christoffel{\alpha}{\mu}{\nu} &= \frac{1}{2} {S^\alpha}_{\mu\nu} - \frac{1}{2} h^\alpha_\sigma {S^\sigma}_{\mu\nu} + \bigo{h^3} , \\
{S^\alpha}_{\mu\nu} &= \partial_\mu h^\alpha_\nu + \partial_\nu h^\alpha_\mu - \partial^\alpha h_{\mu\nu} . \\
\end{split}
\end{equation}
The calculation of the curvature tensors can be done straightforwardly and we obtain
\begin{widetext}
\begin{equation}
\begin{split}
{R^\alpha}_{\beta\gamma\delta} &= \partial_{[\gamma} {S^\alpha}_{\delta]\beta} - h^\alpha_\sigma \partial_{[\gamma} {S^\sigma}_{\delta]\beta} - \frac{1}{2} \eta_{\mu\nu} \eta^{\alpha\sigma} {S^\mu}_{\sigma[\gamma} {S^\nu}_{\delta]\beta} + \bigo{h^3} ,\\
R_{\alpha\beta} &= \frac{1}{2} \left( \partial_\mu {S^\mu}_{\alpha\beta} - \partial_\alpha \partial_\beta h \right) - h^\nu_\mu \partial_{[\nu} {S^\mu}_{\beta]\alpha} - \frac{1}{2} \eta_{\mu\nu} \eta^{\gamma\delta} {S^\nu}_{\delta[\gamma} {S^\mu}_{\beta]\alpha} + \bigo{h^3} ,\\
R &= \left( \partial_\mu \partial_\nu h^{\mu\nu} - \dalembert h \right) + h^{\mu\nu} \left( \partial_\nu \partial_\mu h + \dalembert h_{\mu\nu} - 2 \partial_\nu \partial^\sigma h_{\mu\sigma} \right) \\
&\quad- \frac{1}{4} \left( 2 \partial_\sigma h^{\nu\sigma} - \partial^\nu h \right) \left( 2 \partial^\tau h_{\nu\tau} - \partial_\nu h \right) + \frac{1}{4} \left( 3 \partial_\gamma h_{\mu\delta} - 2 \partial_\mu h_{\gamma\delta} \right) \left( \partial^\gamma h^{\mu\delta} \right) + \bigo{h^3} \eqend{.}
\end{split}
\end{equation}
\end{widetext}
where $\dalembert = \eta^{\mu\nu} \partial_\mu \partial_\nu$ and $\partial^\mu = \eta^{\mu\nu} \partial_\nu$.

The Weyl tensor is given by
\begin{equation}
\label{weyl_tensor_definition}
C^{\alpha\beta}{}_{\gamma\delta} = R^{\alpha\beta}{}_{\gamma\delta} - \frac{4}{(d-2)} R^{[\alpha}_{[\gamma} \delta_{\delta]}^{\beta]} + \frac{2}{(d-1)(d-2)} R \, \delta^\alpha_{[\gamma} \delta_{\delta]}^\beta \eqend{.}
\end{equation}
In four dimensions it obeys the identity
\begin{equation}
\label{fourd_weyl_identity}
C_{\mu\beta\gamma\delta} {C_\nu}^{\beta\gamma\delta} = \frac{1}{4} g_{\mu\nu} C_{\alpha\beta\gamma\delta} C^{\alpha\beta\gamma\delta} \eqend{,}
\end{equation}
which can be proved \cite{lovelock,ddti} by expanding the equality
\begin{equation}
0 = \delta^{[\mu}_{[\nu} {C_{\alpha\beta]}}^{\gamma\delta]} {C^{\alpha\beta}}_{\gamma\delta} \eqend{.}
\end{equation}
Alternatively, it can also be proved using the Gauß-Bonnet theorem.

%% file: appendix_conformal.tex
Under the conformal transformation
\begin{equation}
\tilde{g}_{\mu\nu} = a^2 g_{\mu\nu}
\end{equation}
the Christoffel symbols transform as
\begin{equation}
\varchristoffel{\alpha}{\mu}{\nu} = \christoffel{\alpha}{\mu}{\nu} + a^{-1} \left( \delta^\alpha_\mu \delta^\sigma_\nu + \delta^\alpha_\nu \delta^\sigma_\mu - g_{\mu\nu} g^{\alpha\sigma} \right) \partial_\sigma a \eqend{,}
\end{equation}
and the curvature tensors become
\begin{equation}
\begin{split}
{\tilde{R}^\alpha}{}_{\beta\gamma\delta} &= {R^\alpha}_{\beta\gamma\delta} - 2 a^{-2} \delta^\alpha_{[\gamma} g_{\delta]\beta} ( \nabla^\sigma a ) ( \nabla_\sigma a ) \\
&\quad+ 4 g^{\alpha\tau} \delta^\sigma_{[\gamma} g_{\delta][\tau} \left[ a^{-1} \nabla_{\beta]} \nabla_\sigma a - 2 a^{-2} ( \nabla_{\beta]} a ) ( \nabla_\sigma a ) \right] \\
\tilde{R}_{\mu\nu} &= R_{\mu\nu} - 2 a^{-1} \nabla_\mu \nabla_\nu a + 4 a^{-2} ( \nabla_\mu a ) ( \nabla_\nu a ) \\
&\quad- g_{\mu\nu} \left[ a^{-2} ( \nabla^\sigma a ) ( \nabla_\sigma a ) + a^{-1} \dalembert_g a \right] \\
a^2 \tilde{R} &= R - 6 a^{-1} \dalembert_g a \eqend{,}
\end{split}
\end{equation}
where $\nabla_\mu$ is the covariant derivative associated with 
$g_{\mu\nu}$.

%% file: appendix_special.tex
We define the entire function $\Ein (z)$ by
\begin{equation}
\label{expintegral_ein_definition}
\Ein (z) = \int_0^z \frac{\mathe^t - 1}{t} \total t = \sum_{k=1}^\infty \frac{z^k}{k \, k!} \eqend{.}
\end{equation}
Its asymptotic expansion at infinity (for $\Re \alpha \le 0$) is given by
\begin{equation}
\Ein  (\alpha r ) \sim - \gamma - \ln ( - \alpha r ) + \bigo{\frac{1}{r}} \quad (r \to \infty) \eqend{,}
\end{equation}
where $\gamma$ is the Euler-Mascheroni constant.

For completeness we list here some properties of the Bessel functions which are needed in this paper. Their limits are given by
\begin{align}
J_n(x) &\to \frac{1}{\Gamma(n+1)} \left( \frac{x}{2} \right)^n &(x \to 0)& \label{small_x1} \eqend{,} \\
J_n(x) &\to \sqrt{\frac{2}{\mathpi x}} \cos \left[ x - \frac{\mathpi}{4} (2n + 1) \right] &(x \to \infty)& \label{large_x1} \eqend{,} \\
Y_n(x) &\to - \frac{\Gamma(n)}{\mathpi} \left( \frac{2}{x} \right)^n &(x \to 0)&
\label{small_x2} \eqend{,} \\
Y_n(x) &\to \sqrt{\frac{2}{\mathpi x}} \sin \left[ x - \frac{\mathpi}{4} (2n + 1) \right] &(x \to \infty)& \label{large_x2} \eqend{,}
\end{align}
and for their derivatives we have
\begin{equation}
\begin{split}
\frac{\total}{\total x} \left[ x^n J_n(x) \right] &= x^n J_{n-1}(x) \eqend{,} \\
\frac{\total}{\total x} \left[ x^n Y_n(x) \right] &= x^n Y_{n-1}(x) \eqend{.}
\end{split}
\end{equation}

Expanding the Bessel functions with respect to the order, we get 
\begin{widetext}
\begin{equation}
\begin{split}
\sqrt{\frac{\mathpi}{2}} x^{\frac{3}{2}-\nu} J_{\frac{3}{2}-\nu}(x) &= - x \cos x + \sin x + \nu \bigg[ - 2 \sin x + \left( x \cos x - \sin x \right) \left( \gamma + \ln \left( 2 x^2 \right) \right) \\
&\quad\qquad+ \Im \left[ \Ein\left( 2 \mathi x \right) \left( 1 + \mathi x \right) \mathe^{- \mathi x} \right] \bigg] + \bigo{\nu^2} \label{nu_expansion1} \eqend{,}
\end{split}
\end{equation}
\begin{equation}
\begin{split}
\sqrt{\frac{\mathpi}{2}} x^{\frac{3}{2}-\nu} Y_{\frac{3}{2}-\nu}(x) &= - \cos x - x \sin x + \nu \bigg[ 2 \cos x - \left( \cos x + x \sin x \right) \left( \gamma + \ln 2 \right) \\
&\quad\qquad- \Re \left[ \Ein\left( 2 \mathi x \right) \left( 1 + \mathi x \right) \mathe^{-\mathi x} \right] - \mathpi \left( x \cos x - \sin x \right) \bigg] + \bigo{\nu^2} \label{nu_expansion2} \eqend{.}
\end{split}
\end{equation}
\end{widetext}

%% file: regular_states.tex
In deriving the linearized semiclassical Einstein equation~\eqref{einstein_semiclassical_conformal} from the renormalized CTP effective action \cite{camposverdaguer94,camposverdaguer96}, a number of integration by parts with respect to the spacetime variables $y^\mu$ were performed in order to write the nonlocal term as it appears in eq.~\eqref{einstein_semiclassical_conformal}. After renormalization, the kernel $H(x-x';\bar{\mu})$ in the effective action is a well-defined distribution provided that the metric perturbations fall-off sufficiently fast at spatial infinity and at the asymptotic initial time (there is no such requirement for the asymptotic future due to the causal nature of the kernel). However, when giving initial conditions at some finite time $\eta_0$, the linearized stress tensor expectation value gets boundary contributions from that lower integration limit which diverge when the stress tensor is evaluated at $\eta_0$.

Working in Fourier space for the spatial coordinates, the nonlocal term in eq.~\eqref{einstein_semiclassical_conformal} can be written as
\begin{equation}
\label{nonlocal_expect}
\expect{\op{T}_{\mu\nu}^{(1)}(\eta,\vec{p})}_\text{nl} = \frac{3 \alpha}{a^2(\eta)} \int \tilde{H}(\eta-\eta', \vec{p}; \bar{\mu}) A_{\mu\nu}(\eta',\vec{p}) \total \eta' \eqend{,}
\end{equation}
where the Fourier transformed kernel is given by
\begin{widetext}
\begin{equation}
\label{nonlocal_kernel}
\tilde{H}(\eta-\eta', \vec{p}; \bar{\mu}) = \cos\left[\abs{\vec{p}} (\eta-\eta')\right] \distlim_{\epsilon \to 0} \left[ \frac{\Theta(\eta-\eta'-\epsilon)}{\eta-\eta'} + \delta(\eta-\eta') \left( \ln(\bar{\mu} \epsilon) + \gamma \right) \right] \eqend{;}
\end{equation}
\end{widetext}
see ref.~\cite{frv2011a} for its computation and for further details.
The structure of the nonlocal term for each spatial Fourier mode is then analogous to that for the case of spatially isotropic and homogeneous metric perturbations studied in refs.~\cite{pereznadal08a,pereznadal08b}. There it was shown in detail that boundary terms at $\eta_0$ arise when writing the result in the same form as in eq.~\eqref{nonlocal_expect} and that they give a contribution to $\expect{\op{T}_{\mu\nu}^{(1)}(\eta)}$ which diverges for $\eta = \eta_0$. Furthermore, it was clarified that the reason for such divergences is the fact that although the Bunch-Davies vacuum is a Hadamard state \cite{birrelldavies} with regular UV behavior in de Sitter spacetime, in general that is no longer the case in a perturbed geometry: with respect to well-behaved adiabatic vacua associated with this geometry it exhibits excitations of modes with arbitrarily short wavelengths.

In refs.~\cite{pereznadal08a,pereznadal08b} a simple method was employed for constructing a family of properly ``dressed'' Gaussian initial states which are regular on the perturbed geometry. The states are prepared by evolving an asymptotic Bunch-Davies vacuum state from $-\infty$ to $\eta_0$ in a given (nondynamical) perturbed geometry which is asymptotically de Sitter and matches the dynamical geometry at $\eta_0$. The metric perturbations during this preparation period can be fairly arbitrary, which allows the generation of a wide family of Gaussian states, but need to fulfill a few requirements: they need to fall off sufficiently fast as $\eta \to -\infty$, so that the time integral in the nonlocal term converges, and they need to be small enough so that their contribution to $\expect{\op{T}_{\mu\nu}^{(1)}(\eta)}$ can be treated as a small perturbation.
Moreover, the matching at $\eta_0$ between the nondynamical metric perturbations during the preparation period and the dynamical ones has to be smooth enough: up to fourth order, which is the maximum number of time derivatives that can appear in $A_{\mu\nu}$. In fact, as shown in detail in refs.~\cite{pereznadal08a,pereznadal08b}, requiring a smooth matching up to this order is exactly equivalent to demanding that the states generated in this way are of fourth adiabatic order, the standard requirement for regular states with a finite renormalized stress tensor expectation value \cite{birrelldavies}.

When the preparation method described in the previous paragraph is employed, the nonlocal contribution in eq.~\eqref{nonlocal_expect} can be naturally separated into two contributions which result from the following splitting of the time integral:
\begin{equation}
\label{int_split}
\int_{-\infty}^{\eta} \total \eta' = \int_{\eta_0}^{\eta} \total \eta' + \int_{-\infty}^{\eta_0} \total \eta'
\eqend{.}
\end{equation}
The first term on the right-hand side will contain the dynamical metric perturbations in the integrand and it will vanish when using order reduction since $A_{\mu\nu}$ vanishes in that case, as described in sec.~\ref{linear_nonlocal}. On the other hand, the second term will give a finite contribution in eq.~\eqref{nonlocal_expect}, even for $\eta=\eta_0$, provided that $A_{\mu\nu}$ is regular and vanishes at $\eta=\eta_0$, which follows from the condition of sufficiently smooth matching required above. This term gives a contribution to the right-hand side of eq.~\eqref{einstein_semiclassical_conformal} which can be interpreted as a perturbation of the stress tensor expectation value associated with the modified initial state. In fact, the equation can then be rewritten as eq.~\eqref{semiclassical_inhomogeneous} with
\begin{equation}
\label{st_pert}
\delta T_{\mu\nu}(\eta, \vec{p}) = \frac{3 \alpha}{a^2(\eta)} \int_{-\infty}^{\eta_0} A_{\mu\nu}(\eta',\vec{p}) \, \tilde{H}(\eta-\eta', \vec{p}; \bar{\mu}) \total \eta' \eqend{.}
\end{equation}
The integral is finite for $\eta \geq  \eta_0$ (remember that the metric perturbations are required to fall off sufficiently fast as $\eta \to -\infty$ so that the lower integration limit is convergent) and so is its limit $\eta \to 0$. Therefore, at late times $\delta T_{\mu\nu}$ decays like $1/a^2$, clearly fulfilling the requirement of secs.~\ref{initial_tensor}--\ref{initial_scalar_vector} so that the inhomogeneous solutions of the semiclassical equation do not alter the conclusions about the semiclassical stability of de Sitter spacetime in this context.

%% file: appendix_starobinsky.tex
To our knowledge, the semiclassical equations of motion for the scalar and vector perturbations, eqs.~\eqref{semiclassical_scalar} and \eqref{semiclassical_vector}, have not appeared explicitly in the literature before. On the other hand, eq.~\eqref{semiclassical_tensor} for the tensor perturbations can be directly compared with Starobinsky's eq.~(7) in ref.~\cite{starobinsky81}.
To do so, we need to calculate first the curvature terms for the unperturbed physical metric $g^{(0)}_{\mu\nu} = a^2 \eta_{\mu\nu}$ which were employed by Starobinsky, and which are given by
\begin{equation}
R^0_0 = 3 H^2 \eqend{,} \qquad R = 12 H^2 \eqend{,} \qquad R' = 0 \eqend{.}
\end{equation}
If we insert the following result for the kernel in Fourier space obtained in ref.~\cite{camposverdaguer94}:
\begin{equation}
H(x; \bar{\mu}) = - \frac{1}{2} \int \left[ \ln \abs{\frac{p^2}{\bar{\mu}^2}} - \mathi \mathpi \Theta(-p^2) \sgn p^0 \right] \mathe^{\mathi p x} \frac{\total^4 p}{(2\mathpi)^4} ,
\end{equation}
employ the mode decomposition introduced in eq.~\eqref{tensor_polariz} and divide by $a^2$, eq.~\eqref{semiclassical_tensor} becomes
\begin{widetext}
\begin{equation}
\label{starobinsky_equation}
\begin{split}
&\left[ 1 - \alpha \kappa^2 H^2 - 2 \beta \kappa^2 H^2 \right] \left( g''_\pm + \vec{k}^2 g_\pm \right) + 2 \left[ 1 - \alpha \kappa^2 H^2 - 2 \beta \kappa^2 H^2 \right] H a g'_\pm = \\
&\quad 3 \alpha \kappa^2 a^{-2} \Bigg[ 2 H a \left( g'''_\pm + \vec{k}^2 g'_\pm \right) + H^2 a^2 \left( g''_\pm - \vec{k}^2 g_\pm \right) \\
&\qquad\qquad\qquad- \frac{1}{2} \int \mathe^{- \mathi k^0 \eta} \left[ \int g_\pm(\eta') \mathe^{\mathi k^0 \eta'} \total \eta' \right] (\vec{k}^2)^2 \left[ \ln \abs{\frac{k^2}{\bar{\mu}^2 a^2}} + \mathi \mathpi \Theta(-k^2) \sgn k^0 \right] \frac{\total k^0}{2\mathpi} \Bigg] \eqend{.}
\end{split}
\end{equation}
\end{widetext}
This coincides exactly with Starobinsky's equation if we take into account that
his constants $M^2$, $H^2$ and $G\xi/(60 \mathpi)$ correspond to $2/(\beta \kappa^2)$, $- 2/(\alpha \kappa^2)$ and $3 \alpha \kappa^2$, respectively. Note that Starobinsky's $H^2$ is positive because he considers photon fields, while we have conformal scalars, for which the corresponding constant changes sign \cite{bunchdavies}.

In Starobinsky's inflationary model the expansion is driven by the trace anomaly of the quantized matter fields, so that the actual value of Starobinsky's Hubble parameter $H$ is different from our $H$, where the expansion is driven by the cosmological constant $\Lambda$. Nevertheless, this has no effect on the form of eq.~\eqref{starobinsky_equation}.

%% file: fprv2011.bbl
\begin{thebibliography}{64}%
\makeatletter
\providecommand \@ifxundefined [1]{%
 \@ifx{#1\undefined}
}%
\providecommand \@ifnum [1]{%
 \ifnum #1\expandafter \@firstoftwo
 \else \expandafter \@secondoftwo
 \fi
}%
\providecommand \@ifx [1]{%
 \ifx #1\expandafter \@firstoftwo
 \else \expandafter \@secondoftwo
 \fi
}%
\providecommand \natexlab [1]{#1}%
\providecommand \enquote  [1]{``#1''}%
\providecommand \bibnamefont  [1]{#1}%
\providecommand \bibfnamefont [1]{#1}%
\providecommand \citenamefont [1]{#1}%
\providecommand \href@noop [0]{\@secondoftwo}%
\providecommand \href [0]{\begingroup \@sanitize@url \@href}%
\providecommand \@href[1]{\@@startlink{#1}\@@href}%
\providecommand \@@href[1]{\endgroup#1\@@endlink}%
\providecommand \@sanitize@url [0]{\catcode `\\12\catcode `\$12\catcode
  `\&12\catcode `\#12\catcode `\^12\catcode `\_12\catcode `\%12\relax}%
\providecommand \@@startlink[1]{}%
\providecommand \@@endlink[0]{}%
\providecommand \url  [0]{\begingroup\@sanitize@url \@url }%
\providecommand \@url [1]{\endgroup\@href {#1}{\urlprefix }}%
\providecommand \urlprefix  [0]{URL }%
\providecommand \Eprint [0]{\href }%
\providecommand \doibase [0]{http://dx.doi.org/}%
\providecommand \selectlanguage [0]{\@gobble}%
\providecommand \bibinfo  [0]{\@secondoftwo}%
\providecommand \bibfield  [0]{\@secondoftwo}%
\providecommand \translation [1]{[#1]}%
\providecommand \BibitemOpen [0]{}%
\providecommand \bibitemStop [0]{}%
\providecommand \bibitemNoStop [0]{.\EOS\space}%
\providecommand \EOS [0]{\spacefactor3000\relax}%
\providecommand \BibitemShut  [1]{\csname bibitem#1\endcsname}%
\let\auto@bib@innerbib\@empty
\bibitem [{\citenamefont {Mukhanov}(2005)}]{mukhanov}%
  \BibitemOpen
  \bibfield  {author} {\bibinfo {author} {\bibfnamefont {V.}~\bibnamefont
  {Mukhanov}},\ }\href@noop {} {\emph {\bibinfo {title} {{Physical Foundations
  of Cosmology}}}}\ (\bibinfo  {publisher} {Cambridge University Press},\
  \bibinfo {address} {Cambridge},\ \bibinfo {year} {2005})\BibitemShut
  {NoStop}%
\bibitem [{\citenamefont {Lyth}\ and\ \citenamefont {Liddle}(2009)}]{lyth}%
  \BibitemOpen
  \bibfield  {author} {\bibinfo {author} {\bibfnamefont {D.~H.}\ \bibnamefont
  {Lyth}}\ and\ \bibinfo {author} {\bibfnamefont {A.~R.}\ \bibnamefont
  {Liddle}},\ }\href@noop {} {\emph {\bibinfo {title} {{The primordial density
  perturbation: cosmology, inflation and the origin of structure; rev.
  version}}}}\ (\bibinfo  {publisher} {Cambridge University Press},\ \bibinfo
  {address} {Cambridge},\ \bibinfo {year} {2009})\BibitemShut {NoStop}%
\bibitem [{\citenamefont {Barrow}(1983)}]{barrow83}%
  \BibitemOpen
  \bibfield  {author} {\bibinfo {author} {\bibfnamefont {J.}~\bibnamefont
  {Barrow}},\ }in\ \href@noop {} {\emph {\bibinfo {booktitle} {{The Very Early
  Universe: Proceedings of the Nuffield Workshop}}}}\ (\bibinfo  {publisher}
  {Cambridge University Press},\ \bibinfo {address} {Cambridge},\ \bibinfo
  {year} {1983})\ p.\ \bibinfo {pages} {267}\BibitemShut {NoStop}%
\bibitem [{\citenamefont {Bruni}\ \emph {et~al.}(2002)\citenamefont {Bruni},
  \citenamefont {Mena},\ and\ \citenamefont {Tavakol}}]{bruni01}%
  \BibitemOpen
  \bibfield  {author} {\bibinfo {author} {\bibfnamefont {M.}~\bibnamefont
  {Bruni}}, \bibinfo {author} {\bibfnamefont {F.~C.}\ \bibnamefont {Mena}}, \
  and\ \bibinfo {author} {\bibfnamefont {R.~K.}\ \bibnamefont {Tavakol}},\
  }\href {\doibase 10.1088/0264-9381/19/5/101} {\bibfield  {journal} {\bibinfo
  {journal} {Class. Quant. Grav.}\ }\textbf {\bibinfo {volume} {19}},\ \bibinfo
  {pages} {L23} (\bibinfo {year} {2002})}\BibitemShut {NoStop}%
\bibitem [{\citenamefont {Wald}(1983)}]{wald83}%
  \BibitemOpen
  \bibfield  {author} {\bibinfo {author} {\bibfnamefont {R.~M.}\ \bibnamefont
  {Wald}},\ }\href {\doibase 10.1103/PhysRevD.28.2118} {\bibfield  {journal}
  {\bibinfo  {journal} {Phys. Rev. D}\ }\textbf {\bibinfo {volume} {28}},\
  \bibinfo {pages} {2118} (\bibinfo {year} {1983})}\BibitemShut {NoStop}%
\bibitem [{\citenamefont {Starobinsky}(1983)}]{starobinsky83}%
  \BibitemOpen
  \bibfield  {author} {\bibinfo {author} {\bibfnamefont {A.~A.}\ \bibnamefont
  {Starobinsky}},\ }\href@noop {} {\bibfield  {journal} {\bibinfo  {journal}
  {JETP Lett.}\ }\textbf {\bibinfo {volume} {37}},\ \bibinfo {pages} {66}
  (\bibinfo {year} {1983})}\BibitemShut {NoStop}%
\bibitem [{\citenamefont {Friedrich}(1986)}]{friedrich86}%
  \BibitemOpen
  \bibfield  {author} {\bibinfo {author} {\bibfnamefont {H.}~\bibnamefont
  {Friedrich}},\ }\href {\doibase 10.1007/BF01205488} {\bibfield  {journal}
  {\bibinfo  {journal} {Commun. Math. Phys.}\ }\textbf {\bibinfo {volume}
  {107}},\ \bibinfo {pages} {587} (\bibinfo {year} {1986})}\BibitemShut
  {NoStop}%
\bibitem [{\citenamefont {Anderson}(2005)}]{anderson05}%
  \BibitemOpen
  \bibfield  {author} {\bibinfo {author} {\bibfnamefont {M.~T.}\ \bibnamefont
  {Anderson}},\ }\href {\doibase 10.1007/s00023-005-0224-x} {\bibfield
  {journal} {\bibinfo  {journal} {Ann. Henri Poincar{\'e}}\ }\textbf {\bibinfo
  {volume} {6}},\ \bibinfo {pages} {801} (\bibinfo {year} {2005})}\BibitemShut
  {NoStop}%
\bibitem [{\citenamefont {Birrell}\ and\ \citenamefont
  {Davies}(1984)}]{birrelldavies}%
  \BibitemOpen
  \bibfield  {author} {\bibinfo {author} {\bibfnamefont {N.~D.}\ \bibnamefont
  {Birrell}}\ and\ \bibinfo {author} {\bibfnamefont {P.~C.~W.}\ \bibnamefont
  {Davies}},\ }\href@noop {} {\emph {\bibinfo {title} {{Quantum Fields in
  Curved Space}}}}\ (\bibinfo  {publisher} {Cambridge University Press},\
  \bibinfo {address} {Cambridge},\ \bibinfo {year} {1984})\BibitemShut
  {NoStop}%
\bibitem [{\citenamefont {Wald}(1992)}]{waldqft}%
  \BibitemOpen
  \bibfield  {author} {\bibinfo {author} {\bibfnamefont {R.~M.}\ \bibnamefont
  {Wald}},\ }\href@noop {} {\emph {\bibinfo {title} {{Quantum Field Theory in
  Curved Spacetime and Black Hole Thermodynamics}}}}\ (\bibinfo  {publisher}
  {University of Chicago Press},\ \bibinfo {address} {Chicago},\ \bibinfo
  {year} {1992})\BibitemShut {NoStop}%
\bibitem [{\citenamefont {Marolf}\ and\ \citenamefont
  {Morrison}(2011)}]{marolf10}%
  \BibitemOpen
  \bibfield  {author} {\bibinfo {author} {\bibfnamefont {D.}~\bibnamefont
  {Marolf}}\ and\ \bibinfo {author} {\bibfnamefont {I.~A.}\ \bibnamefont
  {Morrison}},\ }\href {\doibase 10.1103/PhysRevD.84.044040} {\bibfield
  {journal} {\bibinfo  {journal} {Phys. Rev. D}\ }\textbf {\bibinfo {volume}
  {84}},\ \bibinfo {pages} {044040} (\bibinfo {year} {2011})}\BibitemShut
  {NoStop}%
\bibitem [{\citenamefont {Hollands}(2010)}]{hollands10}%
  \BibitemOpen
  \bibfield  {author} {\bibinfo {author} {\bibfnamefont {S.}~\bibnamefont
  {Hollands}},\ }\href@noop {} {\  (\bibinfo {year} {2010})},\ \Eprint
  {http://arxiv.org/abs/1010.5367} {arXiv:1010.5367 [gr-qc]} \BibitemShut
  {NoStop}%
\bibitem [{\citenamefont {Flanagan}\ and\ \citenamefont
  {Wald}(1996)}]{flanaganwald}%
  \BibitemOpen
  \bibfield  {author} {\bibinfo {author} {\bibfnamefont {E.~E.}\ \bibnamefont
  {Flanagan}}\ and\ \bibinfo {author} {\bibfnamefont {R.~M.}\ \bibnamefont
  {Wald}},\ }\href {\doibase 10.1103/PhysRevD.54.6233} {\bibfield  {journal}
  {\bibinfo  {journal} {Phys. Rev. D}\ }\textbf {\bibinfo {volume} {54}},\
  \bibinfo {pages} {6233} (\bibinfo {year} {1996})}\BibitemShut {NoStop}%
\bibitem [{\citenamefont {Anderson}\ \emph {et~al.}(2009)\citenamefont
  {Anderson}, \citenamefont {Molina-Par{\'\i}s},\ and\ \citenamefont
  {Mottola}}]{anderson09}%
  \BibitemOpen
  \bibfield  {author} {\bibinfo {author} {\bibfnamefont {P.~R.}\ \bibnamefont
  {Anderson}}, \bibinfo {author} {\bibfnamefont {C.}~\bibnamefont
  {Molina-Par{\'\i}s}}, \ and\ \bibinfo {author} {\bibfnamefont
  {E.}~\bibnamefont {Mottola}},\ }\href {\doibase 10.1103/PhysRevD.80.084005}
  {\bibfield  {journal} {\bibinfo  {journal} {Phys. Rev. D}\ }\textbf {\bibinfo
  {volume} {80}},\ \bibinfo {pages} {084005} (\bibinfo {year}
  {2009})}\BibitemShut {NoStop}%
\bibitem [{\citenamefont {Ginsparg}\ and\ \citenamefont
  {Perry}(1983)}]{ginsparg83}%
  \BibitemOpen
  \bibfield  {author} {\bibinfo {author} {\bibfnamefont {P.}~\bibnamefont
  {Ginsparg}}\ and\ \bibinfo {author} {\bibfnamefont {M.~J.}\ \bibnamefont
  {Perry}},\ }\href {\doibase 10.1016/0550-3213(83)90636-3} {\bibfield
  {journal} {\bibinfo  {journal} {Nucl. Phys. B}\ }\textbf {\bibinfo {volume}
  {222}},\ \bibinfo {pages} {245} (\bibinfo {year} {1983})}\BibitemShut
  {NoStop}%
\bibitem [{\citenamefont {{Frieman}}\ and\ \citenamefont
  {{Will}}(1982)}]{frieman82}%
  \BibitemOpen
  \bibfield  {author} {\bibinfo {author} {\bibfnamefont {J.~A.}\ \bibnamefont
  {{Frieman}}}\ and\ \bibinfo {author} {\bibfnamefont {C.~M.}\ \bibnamefont
  {{Will}}},\ }\href {\doibase 10.1086/160181} {\bibfield  {journal} {\bibinfo
  {journal} {Astrophys. J.}\ }\textbf {\bibinfo {volume} {259}},\ \bibinfo
  {pages} {437} (\bibinfo {year} {1982})}\BibitemShut {NoStop}%
\bibitem [{\citenamefont {Hsiang}\ \emph {et~al.}(2011)\citenamefont {Hsiang},
  \citenamefont {Ford}, \citenamefont {Lee},\ and\ \citenamefont
  {Yu}}]{fordetal}%
  \BibitemOpen
  \bibfield  {author} {\bibinfo {author} {\bibfnamefont {J.-T.}\ \bibnamefont
  {Hsiang}}, \bibinfo {author} {\bibfnamefont {L.~H.}\ \bibnamefont {Ford}},
  \bibinfo {author} {\bibfnamefont {D.-S.}\ \bibnamefont {Lee}}, \ and\
  \bibinfo {author} {\bibfnamefont {H.-L.}\ \bibnamefont {Yu}},\ }\href
  {\doibase 10.1103/PhysRevD.83.084027} {\bibfield  {journal} {\bibinfo
  {journal} {Phys. Rev. D}\ }\textbf {\bibinfo {volume} {83}},\ \bibinfo
  {pages} {084027} (\bibinfo {year} {2011})}\BibitemShut {NoStop}%
\bibitem [{\citenamefont {Fabris}\ \emph {et~al.}(2012)\citenamefont {Fabris},
  \citenamefont {Pelinson}, \citenamefont {de~O.~Salles},\ and\ \citenamefont
  {Shapiro}}]{pelinson12}%
  \BibitemOpen
  \bibfield  {author} {\bibinfo {author} {\bibfnamefont {J.~C.}\ \bibnamefont
  {Fabris}}, \bibinfo {author} {\bibfnamefont {A.~M.}\ \bibnamefont
  {Pelinson}}, \bibinfo {author} {\bibfnamefont {F.}~\bibnamefont
  {de~O.~Salles}}, \ and\ \bibinfo {author} {\bibfnamefont {I.~L.}\
  \bibnamefont {Shapiro}},\ }\href {\doibase 10.1088/1475-7516/2012/02/019}
  {\bibfield  {journal} {\bibinfo  {journal} {JCAP}\ }\textbf {\bibinfo
  {volume} {2012}},\ \bibinfo {pages} {019} (\bibinfo {year}
  {2012})}\BibitemShut {NoStop}%
\bibitem [{\citenamefont {Starobinsky}(1980)}]{starobinskyinflation}%
  \BibitemOpen
  \bibfield  {author} {\bibinfo {author} {\bibfnamefont {A.~A.}\ \bibnamefont
  {Starobinsky}},\ }\href {\doibase 10.1016/0370-2693(80)90670-X} {\bibfield
  {journal} {\bibinfo  {journal} {Phys. Lett. B}\ }\textbf {\bibinfo {volume}
  {91}},\ \bibinfo {pages} {99} (\bibinfo {year} {1980})}\BibitemShut {NoStop}%
\bibitem [{\citenamefont {Horowitz}(1980)}]{horowitz80}%
  \BibitemOpen
  \bibfield  {author} {\bibinfo {author} {\bibfnamefont {G.~T.}\ \bibnamefont
  {Horowitz}},\ }\href {\doibase 10.1103/PhysRevD.21.1445} {\bibfield
  {journal} {\bibinfo  {journal} {Phys. Rev. D}\ }\textbf {\bibinfo {volume}
  {21}},\ \bibinfo {pages} {1445} (\bibinfo {year} {1980})}\BibitemShut
  {NoStop}%
\bibitem [{\citenamefont {Park}\ and\ \citenamefont {Woodard}(2011)}]{park11}%
  \BibitemOpen
  \bibfield  {author} {\bibinfo {author} {\bibfnamefont {S.}~\bibnamefont
  {Park}}\ and\ \bibinfo {author} {\bibfnamefont {R.}~\bibnamefont {Woodard}},\
  }\href {\doibase 10.1103/PhysRevD.84.124058} {\bibfield  {journal} {\bibinfo
  {journal} {Phys. Rev. D}\ }\textbf {\bibinfo {volume} {84}},\ \bibinfo
  {pages} {124058} (\bibinfo {year} {2011})}\BibitemShut {NoStop}%
\bibitem [{\citenamefont {P{\'e}rez-Nadal}\ \emph
  {et~al.}(2008{\natexlab{a}})\citenamefont {P{\'e}rez-Nadal}, \citenamefont
  {Roura},\ and\ \citenamefont {Verdaguer}}]{pereznadal08a}%
  \BibitemOpen
  \bibfield  {author} {\bibinfo {author} {\bibfnamefont {G.}~\bibnamefont
  {P{\'e}rez-Nadal}}, \bibinfo {author} {\bibfnamefont {A.}~\bibnamefont
  {Roura}}, \ and\ \bibinfo {author} {\bibfnamefont {E.}~\bibnamefont
  {Verdaguer}},\ }\href {\doibase 10.1103/PhysRevD.77.124033} {\bibfield
  {journal} {\bibinfo  {journal} {Phys. Rev. D}\ }\textbf {\bibinfo {volume}
  {77}},\ \bibinfo {pages} {124033} (\bibinfo {year}
  {2008}{\natexlab{a}})}\BibitemShut {NoStop}%
\bibitem [{\citenamefont {P{\'e}rez-Nadal}\ \emph
  {et~al.}(2008{\natexlab{b}})\citenamefont {P{\'e}rez-Nadal}, \citenamefont
  {Roura},\ and\ \citenamefont {Verdaguer}}]{pereznadal08b}%
  \BibitemOpen
  \bibfield  {author} {\bibinfo {author} {\bibfnamefont {G.}~\bibnamefont
  {P{\'e}rez-Nadal}}, \bibinfo {author} {\bibfnamefont {A.}~\bibnamefont
  {Roura}}, \ and\ \bibinfo {author} {\bibfnamefont {E.}~\bibnamefont
  {Verdaguer}},\ }\href {\doibase 10.1088/0264-9381/25/15/154013} {\bibfield
  {journal} {\bibinfo  {journal} {Class. Quant. Grav.}\ }\textbf {\bibinfo
  {volume} {25}},\ \bibinfo {pages} {154013} (\bibinfo {year}
  {2008}{\natexlab{b}})}\BibitemShut {NoStop}%
\bibitem [{\citenamefont {Parker}\ and\ \citenamefont
  {Simon}(1993)}]{simonparker93}%
  \BibitemOpen
  \bibfield  {author} {\bibinfo {author} {\bibfnamefont {L.}~\bibnamefont
  {Parker}}\ and\ \bibinfo {author} {\bibfnamefont {J.~Z.}\ \bibnamefont
  {Simon}},\ }\href {\doibase 10.1103/PhysRevD.47.1339} {\bibfield  {journal}
  {\bibinfo  {journal} {Phys. Rev. D}\ }\textbf {\bibinfo {volume} {47}},\
  \bibinfo {pages} {1339} (\bibinfo {year} {1993})}\BibitemShut {NoStop}%
\bibitem [{\citenamefont {Donoghue}(1994)}]{donoghue94}%
  \BibitemOpen
  \bibfield  {author} {\bibinfo {author} {\bibfnamefont {J.~F.}\ \bibnamefont
  {Donoghue}},\ }\href {\doibase 10.1103/PhysRevD.50.3874} {\bibfield
  {journal} {\bibinfo  {journal} {Phys. Rev. D}\ }\textbf {\bibinfo {volume}
  {50}},\ \bibinfo {pages} {3874} (\bibinfo {year} {1994})}\BibitemShut
  {NoStop}%
\bibitem [{\citenamefont {Burgess}(2004)}]{burgess03}%
  \BibitemOpen
  \bibfield  {author} {\bibinfo {author} {\bibfnamefont {C.}~\bibnamefont
  {Burgess}},\ }\href {http://www.livingreviews.org/lrr-2004-5} {\bibfield
  {journal} {\bibinfo  {journal} {Living Rev. Rel.}\ }\textbf {\bibinfo
  {volume} {7}},\ \bibinfo {pages} {5} (\bibinfo {year} {2004})}\BibitemShut
  {NoStop}%
\bibitem [{\citenamefont {Tsamis}\ and\ \citenamefont
  {Woodard}(1996)}]{tsamis96}%
  \BibitemOpen
  \bibfield  {author} {\bibinfo {author} {\bibfnamefont {N.}~\bibnamefont
  {Tsamis}}\ and\ \bibinfo {author} {\bibfnamefont {R.}~\bibnamefont
  {Woodard}},\ }\href {\doibase 10.1016/0550-3213(96)00246-5} {\bibfield
  {journal} {\bibinfo  {journal} {Nucl. Phys. B}\ }\textbf {\bibinfo {volume}
  {474}},\ \bibinfo {pages} {235} (\bibinfo {year} {1996})}\BibitemShut
  {NoStop}%
\bibitem [{\citenamefont {Tsamis}\ and\ \citenamefont
  {Woodard}(1997)}]{tsamis97}%
  \BibitemOpen
  \bibfield  {author} {\bibinfo {author} {\bibfnamefont {N.}~\bibnamefont
  {Tsamis}}\ and\ \bibinfo {author} {\bibfnamefont {R.}~\bibnamefont
  {Woodard}},\ }\href {\doibase 10.1006/aphy.1997.5613} {\bibfield  {journal}
  {\bibinfo  {journal} {Annals Phys.}\ }\textbf {\bibinfo {volume} {253}},\
  \bibinfo {pages} {1} (\bibinfo {year} {1997})}\BibitemShut {NoStop}%
\bibitem [{\citenamefont {Gerstenlauer}\ \emph {et~al.}(2011)\citenamefont
  {Gerstenlauer}, \citenamefont {Hebecker},\ and\ \citenamefont
  {Tasinato}}]{gerstenlauer11}%
  \BibitemOpen
  \bibfield  {author} {\bibinfo {author} {\bibfnamefont {M.}~\bibnamefont
  {Gerstenlauer}}, \bibinfo {author} {\bibfnamefont {A.}~\bibnamefont
  {Hebecker}}, \ and\ \bibinfo {author} {\bibfnamefont {G.}~\bibnamefont
  {Tasinato}},\ }\href {\doibase 10.1088/1475-7516/2011/06/021} {\bibfield
  {journal} {\bibinfo  {journal} {JCAP}\ }\textbf {\bibinfo {volume} {1106}},\
  \bibinfo {pages} {021} (\bibinfo {year} {2011})}\BibitemShut {NoStop}%
\bibitem [{\citenamefont {Giddings}\ and\ \citenamefont
  {Sloth}(2011)}]{giddings11}%
  \BibitemOpen
  \bibfield  {author} {\bibinfo {author} {\bibfnamefont {S.~B.}\ \bibnamefont
  {Giddings}}\ and\ \bibinfo {author} {\bibfnamefont {M.~S.}\ \bibnamefont
  {Sloth}},\ }\href {\doibase 10.1103/PhysRevD.84.063528} {\bibfield  {journal}
  {\bibinfo  {journal} {Phys. Rev. D}\ }\textbf {\bibinfo {volume} {84}},\
  \bibinfo {pages} {063528} (\bibinfo {year} {2011})}\BibitemShut {NoStop}%
\bibitem [{\citenamefont {Urakawa}\ and\ \citenamefont
  {Tanaka}(2010)}]{urakawa10}%
  \BibitemOpen
  \bibfield  {author} {\bibinfo {author} {\bibfnamefont {Y.}~\bibnamefont
  {Urakawa}}\ and\ \bibinfo {author} {\bibfnamefont {T.}~\bibnamefont
  {Tanaka}},\ }\href {\doibase 10.1103/PhysRevD.82.121301} {\bibfield
  {journal} {\bibinfo  {journal} {Phys. Rev. D}\ }\textbf {\bibinfo {volume}
  {82}},\ \bibinfo {pages} {121301} (\bibinfo {year} {2010})}\BibitemShut
  {NoStop}%
\bibitem [{\citenamefont {P{\'e}rez-Nadal}\ \emph {et~al.}(2010)\citenamefont
  {P{\'e}rez-Nadal}, \citenamefont {Roura},\ and\ \citenamefont
  {Verdaguer}}]{pereznadal10}%
  \BibitemOpen
  \bibfield  {author} {\bibinfo {author} {\bibfnamefont {G.}~\bibnamefont
  {P{\'e}rez-Nadal}}, \bibinfo {author} {\bibfnamefont {A.}~\bibnamefont
  {Roura}}, \ and\ \bibinfo {author} {\bibfnamefont {E.}~\bibnamefont
  {Verdaguer}},\ }\href {\doibase 10.1088/1475-7516/2010/05/036} {\bibfield
  {journal} {\bibinfo  {journal} {JCAP}\ }\textbf {\bibinfo {volume} {1005}},\
  \bibinfo {pages} {036} (\bibinfo {year} {2010})}\BibitemShut {NoStop}%
\bibitem [{\citenamefont {Fr{\"o}b}\ \emph {et~al.}(2012)\citenamefont
  {Fr{\"o}b}, \citenamefont {Roura},\ and\ \citenamefont
  {Verdaguer}}]{frv2011a}%
  \BibitemOpen
  \bibfield  {author} {\bibinfo {author} {\bibfnamefont {M.~B.}\ \bibnamefont
  {Fr{\"o}b}}, \bibinfo {author} {\bibfnamefont {A.}~\bibnamefont {Roura}}, \
  and\ \bibinfo {author} {\bibfnamefont {E.}~\bibnamefont {Verdaguer}},\ }\href
  {\doibase 10.1088/1475-7516/2012/08/009} {\bibfield  {journal} {\bibinfo
  {journal} {JCAP}\ }\textbf {\bibinfo {volume} {1208}},\ \bibinfo {pages}
  {009} (\bibinfo {year} {2012})}\BibitemShut {NoStop}%
\bibitem [{\citenamefont {Fr{\"o}b}\ \emph {et~al.}(2013)\citenamefont
  {Fr{\"o}b}, \citenamefont {Roura},\ and\ \citenamefont
  {Verdaguer}}]{frv2012}%
  \BibitemOpen
  \bibfield  {author} {\bibinfo {author} {\bibfnamefont {M.~B.}\ \bibnamefont
  {Fr{\"o}b}}, \bibinfo {author} {\bibfnamefont {A.}~\bibnamefont {Roura}}, \
  and\ \bibinfo {author} {\bibfnamefont {E.}~\bibnamefont {Verdaguer}},\
  }\href@noop {} {\enquote {\bibinfo {title} {{The one-loop Riemann tensor
  correlator in de Sitter spacetime. Conformal case.}}}\ } (\bibinfo {year}
  {2013}),\ \bibinfo {note} {in preparation}\BibitemShut {NoStop}%
\bibitem [{\citenamefont {Starobinski}(1981)}]{starobinsky81}%
  \BibitemOpen
  \bibfield  {author} {\bibinfo {author} {\bibfnamefont {A.~A.}\ \bibnamefont
  {Starobinski}},\ }\href@noop {} {\bibfield  {journal} {\bibinfo  {journal}
  {JETP Letters}\ }\textbf {\bibinfo {volume} {34}},\ \bibinfo {pages} {438}
  (\bibinfo {year} {1981})}\BibitemShut {NoStop}%
\bibitem [{\citenamefont {Misner}\ \emph {et~al.}(1973)\citenamefont {Misner},
  \citenamefont {Thorne},\ and\ \citenamefont {Wheeler}}]{mtw}%
  \BibitemOpen
  \bibfield  {author} {\bibinfo {author} {\bibfnamefont {C.}~\bibnamefont
  {Misner}}, \bibinfo {author} {\bibfnamefont {K.}~\bibnamefont {Thorne}}, \
  and\ \bibinfo {author} {\bibfnamefont {J.}~\bibnamefont {Wheeler}},\
  }\href@noop {} {\emph {\bibinfo {title} {{Gravitation}}}}\ (\bibinfo
  {publisher} {W. H. Freeman},\ \bibinfo {year} {1973})\BibitemShut {NoStop}%
\bibitem [{\citenamefont {Jackson}(1998)}]{jackson}%
  \BibitemOpen
  \bibfield  {author} {\bibinfo {author} {\bibfnamefont {J.~D.}\ \bibnamefont
  {Jackson}},\ }\href@noop {} {\emph {\bibinfo {title} {{Classical
  Electrodynamics}}}},\ \bibinfo {edition} {3rd}\ ed.\ (\bibinfo  {publisher}
  {Wiley},\ \bibinfo {year} {1998})\BibitemShut {NoStop}%
\bibitem [{\citenamefont {Landau}\ and\ \citenamefont
  {Lifshitz}(1962)}]{lifshitzbook}%
  \BibitemOpen
  \bibfield  {author} {\bibinfo {author} {\bibfnamefont {L.~D.}\ \bibnamefont
  {Landau}}\ and\ \bibinfo {author} {\bibfnamefont {E.~M.}\ \bibnamefont
  {Lifshitz}},\ }\href@noop {} {\emph {\bibinfo {title} {{The classical theory
  of fields. Translated from the Russian by Morton Hamermesh}}}},\ \bibinfo
  {edition} {revised 2nd}\ ed.\ (\bibinfo  {publisher} {Pergamon Press},\
  \bibinfo {year} {1962})\BibitemShut {NoStop}%
\bibitem [{\citenamefont {Bel}\ \emph {et~al.}(1981)\citenamefont {Bel},
  \citenamefont {Damour}, \citenamefont {Deruelle}, \citenamefont {Ibanez},\
  and\ \citenamefont {Martin}}]{beletal81}%
  \BibitemOpen
  \bibfield  {author} {\bibinfo {author} {\bibfnamefont {L.}~\bibnamefont
  {Bel}}, \bibinfo {author} {\bibfnamefont {T.}~\bibnamefont {Damour}},
  \bibinfo {author} {\bibfnamefont {N.}~\bibnamefont {Deruelle}}, \bibinfo
  {author} {\bibfnamefont {J.}~\bibnamefont {Ibanez}}, \ and\ \bibinfo {author}
  {\bibfnamefont {J.}~\bibnamefont {Martin}},\ }\href {\doibase
  10.1007/BF00756073} {\bibfield  {journal} {\bibinfo  {journal} {Gen. Rel.
  Grav.}\ }\textbf {\bibinfo {volume} {13}},\ \bibinfo {pages} {963} (\bibinfo
  {year} {1981})}\BibitemShut {NoStop}%
\bibitem [{\citenamefont {Bel}\ and\ \citenamefont {Zia}(1985)}]{belzia85}%
  \BibitemOpen
  \bibfield  {author} {\bibinfo {author} {\bibfnamefont {L.}~\bibnamefont
  {Bel}}\ and\ \bibinfo {author} {\bibfnamefont {H.~S.}\ \bibnamefont {Zia}},\
  }\href {\doibase 10.1103/PhysRevD.32.3128} {\bibfield  {journal} {\bibinfo
  {journal} {Phys. Rev. D}\ }\textbf {\bibinfo {volume} {32}},\ \bibinfo
  {pages} {3128} (\bibinfo {year} {1985})}\BibitemShut {NoStop}%
\bibitem [{\citenamefont {Simon}(1992)}]{simon92}%
  \BibitemOpen
  \bibfield  {author} {\bibinfo {author} {\bibfnamefont {J.~Z.}\ \bibnamefont
  {Simon}},\ }\href {\doibase 10.1103/PhysRevD.45.1953} {\bibfield  {journal}
  {\bibinfo  {journal} {Phys. Rev. D}\ }\textbf {\bibinfo {volume} {45}},\
  \bibinfo {pages} {1953} (\bibinfo {year} {1992})}\BibitemShut {NoStop}%
\bibitem [{\citenamefont {Campos}\ and\ \citenamefont
  {Verdaguer}(1994)}]{camposverdaguer94}%
  \BibitemOpen
  \bibfield  {author} {\bibinfo {author} {\bibfnamefont {A.}~\bibnamefont
  {Campos}}\ and\ \bibinfo {author} {\bibfnamefont {E.}~\bibnamefont
  {Verdaguer}},\ }\href {\doibase 10.1103/PhysRevD.49.1861} {\bibfield
  {journal} {\bibinfo  {journal} {Phys. Rev. D}\ }\textbf {\bibinfo {volume}
  {49}},\ \bibinfo {pages} {1861} (\bibinfo {year} {1994})}\BibitemShut
  {NoStop}%
\bibitem [{\citenamefont {Campos}\ and\ \citenamefont
  {Verdaguer}(1996)}]{camposverdaguer96}%
  \BibitemOpen
  \bibfield  {author} {\bibinfo {author} {\bibfnamefont {A.}~\bibnamefont
  {Campos}}\ and\ \bibinfo {author} {\bibfnamefont {E.}~\bibnamefont
  {Verdaguer}},\ }\href {\doibase 10.1103/PhysRevD.53.1927} {\bibfield
  {journal} {\bibinfo  {journal} {Phys. Rev. D}\ }\textbf {\bibinfo {volume}
  {53}},\ \bibinfo {pages} {1927} (\bibinfo {year} {1996})}\BibitemShut
  {NoStop}%
\bibitem [{\citenamefont {Horowitz}\ and\ \citenamefont
  {Wald}(1980)}]{horowitzwald1}%
  \BibitemOpen
  \bibfield  {author} {\bibinfo {author} {\bibfnamefont {G.~T.}\ \bibnamefont
  {Horowitz}}\ and\ \bibinfo {author} {\bibfnamefont {R.~M.}\ \bibnamefont
  {Wald}},\ }\href {\doibase 10.1103/PhysRevD.21.1462} {\bibfield  {journal}
  {\bibinfo  {journal} {Phys. Rev. D}\ }\textbf {\bibinfo {volume} {21}},\
  \bibinfo {pages} {1462} (\bibinfo {year} {1980})}\BibitemShut {NoStop}%
\bibitem [{\citenamefont {Horowitz}\ and\ \citenamefont
  {Wald}(1982)}]{horowitzwald2}%
  \BibitemOpen
  \bibfield  {author} {\bibinfo {author} {\bibfnamefont {G.~T.}\ \bibnamefont
  {Horowitz}}\ and\ \bibinfo {author} {\bibfnamefont {R.~M.}\ \bibnamefont
  {Wald}},\ }\href {\doibase 10.1103/PhysRevD.25.3408} {\bibfield  {journal}
  {\bibinfo  {journal} {Phys. Rev. D}\ }\textbf {\bibinfo {volume} {25}},\
  \bibinfo {pages} {3408} (\bibinfo {year} {1982})}\BibitemShut {NoStop}%
\bibitem [{\citenamefont {Dowker}\ and\ \citenamefont
  {Critchley}(1976)}]{dowker76}%
  \BibitemOpen
  \bibfield  {author} {\bibinfo {author} {\bibfnamefont {J.~S.}\ \bibnamefont
  {Dowker}}\ and\ \bibinfo {author} {\bibfnamefont {R.}~\bibnamefont
  {Critchley}},\ }\href {\doibase 10.1103/PhysRevD.13.3224} {\bibfield
  {journal} {\bibinfo  {journal} {Phys. Rev. D}\ }\textbf {\bibinfo {volume}
  {13}},\ \bibinfo {pages} {3224} (\bibinfo {year} {1976})}\BibitemShut
  {NoStop}%
\bibitem [{\citenamefont {Wada}\ and\ \citenamefont {Azuma}(1983)}]{wada83}%
  \BibitemOpen
  \bibfield  {author} {\bibinfo {author} {\bibfnamefont {S.}~\bibnamefont
  {Wada}}\ and\ \bibinfo {author} {\bibfnamefont {T.}~\bibnamefont {Azuma}},\
  }\href {\doibase 10.1016/0370-2693(83)90315-5} {\bibfield  {journal}
  {\bibinfo  {journal} {Phys. Lett. B}\ }\textbf {\bibinfo {volume} {132}},\
  \bibinfo {pages} {313} (\bibinfo {year} {1983})}\BibitemShut {NoStop}%
\bibitem [{\citenamefont {Vilenkin}(1985)}]{vilenkin85}%
  \BibitemOpen
  \bibfield  {author} {\bibinfo {author} {\bibfnamefont {A.}~\bibnamefont
  {Vilenkin}},\ }\href {\doibase 10.1103/PhysRevD.32.2511} {\bibfield
  {journal} {\bibinfo  {journal} {Phys. Rev. D}\ }\textbf {\bibinfo {volume}
  {32}},\ \bibinfo {pages} {2511} (\bibinfo {year} {1985})}\BibitemShut
  {NoStop}%
\bibitem [{\citenamefont {Lifshitz}(1946)}]{lifshitz}%
  \BibitemOpen
  \bibfield  {author} {\bibinfo {author} {\bibfnamefont {E.}~\bibnamefont
  {Lifshitz}},\ }\href@noop {} {\bibfield  {journal} {\bibinfo  {journal} {J.
  Phys. (USSR)}\ }\textbf {\bibinfo {volume} {10}},\ \bibinfo {pages} {116}
  (\bibinfo {year} {1946})}\BibitemShut {NoStop}%
\bibitem [{\citenamefont {Bardeen}(1980)}]{bardeen}%
  \BibitemOpen
  \bibfield  {author} {\bibinfo {author} {\bibfnamefont {J.~M.}\ \bibnamefont
  {Bardeen}},\ }\href {\doibase 10.1103/PhysRevD.22.1882} {\bibfield  {journal}
  {\bibinfo  {journal} {Phys. Rev. D}\ }\textbf {\bibinfo {volume} {22}},\
  \bibinfo {pages} {1882} (\bibinfo {year} {1980})}\BibitemShut {NoStop}%
\bibitem [{\citenamefont {Stewart}(1990)}]{stewart90}%
  \BibitemOpen
  \bibfield  {author} {\bibinfo {author} {\bibfnamefont {J.}~\bibnamefont
  {Stewart}},\ }\href@noop {} {\bibfield  {journal} {\bibinfo  {journal}
  {Class. Quant. Grav.}\ }\textbf {\bibinfo {volume} {7}},\ \bibinfo {pages}
  {1169} (\bibinfo {year} {1990})}\BibitemShut {NoStop}%
\bibitem [{\citenamefont {Brill}\ and\ \citenamefont {Hartle}(1964)}]{brill64}%
  \BibitemOpen
  \bibfield  {author} {\bibinfo {author} {\bibfnamefont {D.~R.}\ \bibnamefont
  {Brill}}\ and\ \bibinfo {author} {\bibfnamefont {J.~B.}\ \bibnamefont
  {Hartle}},\ }\href {\doibase 10.1103/PhysRev.135.B271} {\bibfield  {journal}
  {\bibinfo  {journal} {Phys. Rev.}\ }\textbf {\bibinfo {volume} {135}},\
  \bibinfo {pages} {B271} (\bibinfo {year} {1964})}\BibitemShut {NoStop}%
\bibitem [{\citenamefont {Isaacson}(1968)}]{isaacson68}%
  \BibitemOpen
  \bibfield  {author} {\bibinfo {author} {\bibfnamefont {R.~A.}\ \bibnamefont
  {Isaacson}},\ }\href {\doibase 10.1103/PhysRev.166.1272} {\bibfield
  {journal} {\bibinfo  {journal} {Phys. Rev.}\ }\textbf {\bibinfo {volume}
  {166}},\ \bibinfo {pages} {1272} (\bibinfo {year} {1968})}\BibitemShut
  {NoStop}%
\bibitem [{\citenamefont {Martin}\ and\ \citenamefont
  {Verdaguer}(1999)}]{martin99}%
  \BibitemOpen
  \bibfield  {author} {\bibinfo {author} {\bibfnamefont {R.}~\bibnamefont
  {Martin}}\ and\ \bibinfo {author} {\bibfnamefont {E.}~\bibnamefont
  {Verdaguer}},\ }\href {\doibase 10.1103/PhysRevD.60.084008} {\bibfield
  {journal} {\bibinfo  {journal} {Phys. Rev. D}\ }\textbf {\bibinfo {volume}
  {60}},\ \bibinfo {pages} {084008} (\bibinfo {year} {1999})}\BibitemShut
  {NoStop}%
\bibitem [{\citenamefont {Mottola}(1986)}]{mottola86}%
  \BibitemOpen
  \bibfield  {author} {\bibinfo {author} {\bibfnamefont {E.}~\bibnamefont
  {Mottola}},\ }\href {\doibase 10.1103/PhysRevD.33.2136} {\bibfield  {journal}
  {\bibinfo  {journal} {Phys. Rev. D}\ }\textbf {\bibinfo {volume} {33}},\
  \bibinfo {pages} {2136} (\bibinfo {year} {1986})}\BibitemShut {NoStop}%
\bibitem [{\citenamefont {Roura}\ and\ \citenamefont
  {Verdaguer}(2008)}]{roura08}%
  \BibitemOpen
  \bibfield  {author} {\bibinfo {author} {\bibfnamefont {A.}~\bibnamefont
  {Roura}}\ and\ \bibinfo {author} {\bibfnamefont {E.}~\bibnamefont
  {Verdaguer}},\ }\href {\doibase 10.1103/PhysRevD.78.064010} {\bibfield
  {journal} {\bibinfo  {journal} {Phys. Rev. D}\ }\textbf {\bibinfo {volume}
  {78}},\ \bibinfo {pages} {064010} (\bibinfo {year} {2008})}\BibitemShut
  {NoStop}%
\bibitem [{\citenamefont {Mazur}\ and\ \citenamefont
  {Mottola}(2001)}]{mazur01}%
  \BibitemOpen
  \bibfield  {author} {\bibinfo {author} {\bibfnamefont {P.~O.}\ \bibnamefont
  {Mazur}}\ and\ \bibinfo {author} {\bibfnamefont {E.}~\bibnamefont
  {Mottola}},\ }\href {\doibase 10.1103/PhysRevD.64.104022} {\bibfield
  {journal} {\bibinfo  {journal} {Phys. Rev. D}\ }\textbf {\bibinfo {volume}
  {64}},\ \bibinfo {pages} {104022} (\bibinfo {year} {2001})}\BibitemShut
  {NoStop}%
\bibitem [{\citenamefont {Eftekharzadeh}\ \emph {et~al.}(2012)\citenamefont
  {Eftekharzadeh}, \citenamefont {Bates}, \citenamefont {Roura}, \citenamefont
  {Anderson},\ and\ \citenamefont {Hu}}]{eftekharzadeh12}%
  \BibitemOpen
  \bibfield  {author} {\bibinfo {author} {\bibfnamefont {A.}~\bibnamefont
  {Eftekharzadeh}}, \bibinfo {author} {\bibfnamefont {J.~D.}\ \bibnamefont
  {Bates}}, \bibinfo {author} {\bibfnamefont {A.}~\bibnamefont {Roura}},
  \bibinfo {author} {\bibfnamefont {P.~R.}\ \bibnamefont {Anderson}}, \ and\
  \bibinfo {author} {\bibfnamefont {B.}~\bibnamefont {Hu}},\ }\href {\doibase
  10.1103/PhysRevD.85.044037} {\bibfield  {journal} {\bibinfo  {journal} {Phys.
  Rev. D}\ }\textbf {\bibinfo {volume} {85}},\ \bibinfo {pages} {044037}
  (\bibinfo {year} {2012})}\BibitemShut {NoStop}%
\bibitem [{\citenamefont {Osborn}\ and\ \citenamefont
  {Petkou}(1994)}]{osborn94}%
  \BibitemOpen
  \bibfield  {author} {\bibinfo {author} {\bibfnamefont {H.}~\bibnamefont
  {Osborn}}\ and\ \bibinfo {author} {\bibfnamefont {A.}~\bibnamefont
  {Petkou}},\ }\href {\doibase 10.1006/aphy.1994.1045} {\bibfield  {journal}
  {\bibinfo  {journal} {Annals Phys.}\ }\textbf {\bibinfo {volume} {231}},\
  \bibinfo {pages} {311} (\bibinfo {year} {1994})}\BibitemShut {NoStop}%
\bibitem [{\citenamefont {Hu}\ and\ \citenamefont
  {Verdaguer}(2008)}]{stochasticgravity}%
  \BibitemOpen
  \bibfield  {author} {\bibinfo {author} {\bibfnamefont {B.~L.}\ \bibnamefont
  {Hu}}\ and\ \bibinfo {author} {\bibfnamefont {E.}~\bibnamefont {Verdaguer}},\
  }\href {http://www.livingreviews.org/lrr-2008-3} {\bibfield  {journal}
  {\bibinfo  {journal} {Liv. Rev. Rel.}\ } (\bibinfo {year}
  {2008})}\BibitemShut {NoStop}%
\bibitem [{\citenamefont {Hu}\ \emph {et~al.}(2004)\citenamefont {Hu},
  \citenamefont {Roura},\ and\ \citenamefont {Verdaguer}}]{hu04}%
  \BibitemOpen
  \bibfield  {author} {\bibinfo {author} {\bibfnamefont {B.}~\bibnamefont
  {Hu}}, \bibinfo {author} {\bibfnamefont {A.}~\bibnamefont {Roura}}, \ and\
  \bibinfo {author} {\bibfnamefont {E.}~\bibnamefont {Verdaguer}},\ }\href
  {\doibase 10.1103/PhysRevD.70.044002} {\bibfield  {journal} {\bibinfo
  {journal} {Phys. Rev. D}\ }\textbf {\bibinfo {volume} {70}},\ \bibinfo
  {pages} {044002} (\bibinfo {year} {2004})}\BibitemShut {NoStop}%
\bibitem [{\citenamefont {Lovelock}(1970)}]{lovelock}%
  \BibitemOpen
  \bibfield  {author} {\bibinfo {author} {\bibfnamefont {D.}~\bibnamefont
  {Lovelock}},\ }\href {\doibase 10.1017/S0305004100046144} {\bibfield
  {journal} {\bibinfo  {journal} {Math. Proc. Cambr. Phil. Soc.}\ }\textbf
  {\bibinfo {volume} {68}},\ \bibinfo {pages} {345} (\bibinfo {year}
  {1970})}\BibitemShut {NoStop}%
\bibitem [{\citenamefont {Edgar}\ and\ \citenamefont
  {H{\"o}glund}(2002)}]{ddti}%
  \BibitemOpen
  \bibfield  {author} {\bibinfo {author} {\bibfnamefont {S.}~\bibnamefont
  {Edgar}}\ and\ \bibinfo {author} {\bibfnamefont {A.}~\bibnamefont
  {H{\"o}glund}},\ }\href {\doibase 10.1063/1.1425428} {\bibfield  {journal}
  {\bibinfo  {journal} {J. Math. Phys.}\ }\textbf {\bibinfo {volume} {43}},\
  \bibinfo {pages} {659} (\bibinfo {year} {2002})}\BibitemShut {NoStop}%
\bibitem [{\citenamefont {Bunch}\ and\ \citenamefont
  {Davies}(1977)}]{bunchdavies}%
  \BibitemOpen
  \bibfield  {author} {\bibinfo {author} {\bibfnamefont {T.~S.}\ \bibnamefont
  {Bunch}}\ and\ \bibinfo {author} {\bibfnamefont {P.~C.~W.}\ \bibnamefont
  {Davies}},\ }\href {\doibase 10.1098/rspa.1977.0151} {\bibfield  {journal}
  {\bibinfo  {journal} {Proc. R. Soc. A}\ }\textbf {\bibinfo {volume} {356}},\
  \bibinfo {pages} {569} (\bibinfo {year} {1977})}\BibitemShut {NoStop}%
\end{thebibliography}%
